\documentclass[useAMS,usenatbib,referee]{mn2e}
\usepackage{amssymb}
\usepackage{graphicx}
\usepackage{threeparttablex}
\usepackage{longtable} 
\usepackage{fixltx2e}
\usepackage{lscape}
\title[Multi-wavelength Survey of Cluster AGN]{A Multi-wavelength Survey of AGN in Massive Clusters: AGN Detection and Cluster AGN Fraction}
\author[Alison J. Klesman and Vicki L. Sarajedini]{Alison J. Klesman$^{1}$\thanks{E-mail:
alichan@astro.ufl.edu (AJK); vicki@astro.ufl.edu (VLS)} and Vicki L. Sarajedini$^{1}$\\
$^{1}$University of Florida, Department of Astronomy, 211 Bryant Space Science Center, FL, 32611 USA}
\begin{document}

\maketitle

\label{firstpage}

\begin{abstract}
We aim to study the effect of environment on the presence and fuelling of Active Galactic Nuclei (AGN) in massive galaxy clusters. We explore the use of different AGN detection techniques with the goal of selecting AGN across a broad range of luminosities, AGN/host galaxy flux ratios, and obscuration levels. From a sample of 12 galaxy clusters at redshifts 0.5 $<$ z $<$ 0.9, we identify AGN candidates using optical variability from multi-epoch HST imaging, X-ray point sources in Chandra images, and mid-IR SED power-law fits through the Spitzer IRAC channels. We find 178 optical variables, 74 X-ray point sources, and 64 IR power law sources, resulting in an average of $\sim$25 AGN per cluster. We find no significant difference between the fraction of AGN among galaxies in clusters and the percentage of similarly-detected AGN in field galaxy studies ($\sim$2.5\%). This result provides evidence that galaxies are still able to fuel accretion onto their supermassive black holes, even in dense environments. We also investigate correlations between the percentage of AGN and cluster physical properties such as mass, X-ray luminosity, size, morphology class and redshift. We find no significant correlations among cluster properties and the percentage of AGN detected.
\end{abstract}


\section{Introduction}

Active Galactic Nuclei (AGN) are envisioned to be accreting supermassive black holes at the centres of galaxies, with masses ranging from $\sim$10$^{6}$--$10^{10} M_{\odot}$ \citep[e.g.,][]{salpeter64, lb69}. Supermassive black holes are now believed to reside in the centre of all massive galaxies \citep{kr95}, and there is a well-established correlation between the mass of the central black hole and the mass and velocity dispersion of the host galaxyÕs spheroidal component \citep{mag98, fm00, geb00}. This relationship indicates that there is a fundamental link between the growth of supermassive black holes and the formation of the galaxies in which they reside. The first challenge to understanding this link, is the unbiased identification of AGN in galaxy surveys. While AGN are identified using a range of techniques (emission line characteristics, optical and IR colour selection, X-ray emission, radio emission, and variability), many produce an incomplete picture of AGN, as they can be biased against galaxies in which the AGN light is not a significant percentage of the total galaxy light. Optical and UV surveys can also miss more heavily obscured AGN and host galaxies. Therefore, it is necessary to utilise more than one technique to select a more complete sample of AGN to study the links between these objects and the galaxies in which they reside.

AGN show variability on a variety of timescales. The optical line emission and continuum flux from quasars has been observed to vary on timescales of months to years, while X-ray flux from these sources varies on shorter timescales of hours to days (\citealt{peterson01} \& references therein). It has been shown that 80--100\% of AGN candidates display variability up to several percent in the optical over the course of several years \citep[e.g.,][]{kkc86, mcleod2010}. The use of small aperture photometry with high-resolution HST images allows for the detection of varying AGN comprising as little as $\sim$5\% of the total optical galaxy light \citep{sarajedini00, sarajedini03, sarajedini06}. \citet{ks07} investigate the use of multi-wavelength AGN identification techniques by conducting a variability study on a pre-selected sample of X-ray and mid-IR AGN in the Great Observatories Origins Deep Survey (GOODS) South field. They found that 26\% of all AGN candidates (either X-ray or mid-IR selected) are optically varying on timescales of several months. The fraction of optical variables increases to 51\% when considering sources displaying softer X-ray spectra. While unobscured AGN appear to have the most significant optical variability, some obscured AGN were also observed as optical variables. The overall survey results revealed that 2\% of all galaxies in the GOODS North and South fields appear to host varying AGN (Sarajedini et al. 2011).

AGN are also known to be luminous X-ray sources and can be routinely selected from deep X-ray ($<$10 keV) surveys. The X-ray emission is believed to originate very close to the central black hole and is produced by Compton upscattering of softer photons by a hot ``corona'' around the accretion disk \citep[e.g.,][]{st80}. X-ray, UV, and optical emission can be absorbed by dust surrounding the nucleus, and then reprocessed and re-radiated as infrared light. \citet{donley08} find that a power-law fit to photometric measurements through the mid-IR bands can also be used to identify AGN and does so more reliably than using IR colours alone (e.g. Lacy et al. 2004).

Galaxy clusters present a unique environment in which to explore the link between AGN and their host galaxies. The dense cluster environment is known to impact galaxy morphology since a larger fraction of early-type galaxies are found among clusters than that found among field galaxies \citep[e.g.,][]{hh31, morgan61, abell65, oemler74}. However, galaxies in clusters encounter each other at higher relative speeds, resulting in fewer mergers among cluster galaxies than in more loosely bound groups \citep[e.g.,][]{mh97}. Early spectroscopic studies \citep{gisler78, dressler85} presented evidence that AGN may be less common in rich clusters than in the field. These studies concluded that the fraction of AGN in clusters is about 1\%, compared with 5\% observed in the field \citep{dressler85}. This view has been challenged by recent X-ray surveys which have detected large numbers of optically normal cluster galaxies whose luminous X-ray emission would indicate AGN activity \citep[e.g.,][]{martini02, martini06}. Using X-ray observations, \citet{martini02} found a cluster AGN fraction approximately 5 times higher than that found by \citet{dressler85}. To fully address the AGN fraction in clusters and understand the role environment plays requires cluster surveys that identify AGN having a broad range of luminosities and obscuration levels.

In this paper, we will explore the issue of AGN identification in cluster galaxies using a multi-wavelength, multi-technique approach to produce a largely unbiased sample with which to investigate the relationship between AGN, their environment, and their host galaxies. We present our sample of 12 massive galaxy clusters at z = 0.5--0.9 in Section~\ref{clustersample}. In Sections~\ref{optvar},~\ref{xdet}, and~\ref{irdet}, we apply the techniques of optical variability, X-ray emission, and IR power-law SED fitting to identify AGN in these clusters. In Section~\ref{overlaps}, we compare the various identification techniques to explore completeness issues. We determine cluster membership probabilities for the galaxies in our images and calculate the percentage of AGN in clusters in Sections~\ref{globalenvironment} and~\ref{AGNdiscussion}. Finally, we present our conclusions in Section~\ref{conclusions}. Throughout the paper we assume a standard cosmology with H$_{0}$ = 70, $\Omega_{M}$ = 0.3, and $\Omega_{\Lambda}$ = 0.7.

\section{Galaxy Cluster Sample}
\label{clustersample}

To investigate the role of environment on the AGN phenomenon, we selected several galaxy clusters to perform a census of AGN. We chose clusters for which archival data was available that would allow AGN to be detected via optical variability, X-ray emission, and mid-Infrared properties. All of the galaxy clusters used in this work are part of a sample observed multiple times with the Hubble Space Telescope (HST) Advanced Camera for Surveys (ACS) in a survey for supernovae in massive high-redshift clusters (HST programs GO 10493, cycle 14 and GO 10793, cycle 15, P.I. Avishay Gal-Yam). Our sample consists of 12 clusters with redshifts ranging from 0.50--0.89. All have also been observed with the Chandra X-ray observatory, and seven have been imaged in the mid-IR with the Spitzer IRAC instrument. Table~\ref{clustersgen} lists the clusters, their redshifts, and the number of optical ACS epochs from the survey \citep[Table 1]{sharon2010}. The table also lists whether Spitzer data are available for the clusters. 

\subsection{MACS Clusters}

Seven of the clusters in this work are part of the Massive Cluster Survey (MACS) \citep{Ebeling01}. This survey targets distant (z $>$ 0.3), luminous (L$_{X} > 10^{44}$ erg/s), and therefore massive galaxy clusters selected from the ROSAT All-Sky Survey. Detailed optical and X-ray observations were carried out by \citet{barrett06}, including spectroscopic confirmation of cluster members, mass, velocity dispersion, and virial radius determinations. \citet{stott07} report the X-ray luminosities for these clusters. These values are also listed in Table~\ref{clusterprops}.

We note that the virial masses of the MACS clusters found in Table~\ref{clusterprops} are actually measures of M$_{X,200}$ derived by \citet{barrett06} from the X-ray emitting gas mass within the virial radius r$_{v}$ assuming the gas is a perfect gas in hydrostatic equilibrium, isothermal, and spherically symmetric. This mass was used as a proxy for the virial mass determined using the position and velocity dispersion of cluster galaxies, as measurements of M$_{200}$ using this method require a large number of galaxy redshifts, which is difficult and time-consuming, especially for more distant clusters. \citet{barrett06} examines the correlation between cluster virial mass measured via these two methods, using an average of $\sim$40 galaxy redshifts per cluster, and finds the two agree to within 15\%. This difference appears to arise from poor sampling of galaxies when determining the virial mass, and is similar to the findings of \citet{rb02}, who find an agreement of $\sim$25\% between virial masses determined via X-ray gas and galaxy dynamics with a similar sampling of $\sim$40 galaxies per cluster. \citet{rines03} find that the virial mass determination given by these two methods agrees well when using a sample of $\sim$200 galaxies to determine the virial mass via galaxy dynamics.

\subsection{CL0152-1357}

CL0152 has a bolometric X-ray luminosity of L$_{X}$ = 1.6 x 10$^{45}$ erg/s \citep{maughan03} and a virial radius r$_{200}$ of 1.14 Mpc \citep{jee05}. The cluster shows significant substructure in its X-ray morphology indicative of an ongoing merger, with two peaks in its X-ray morphology that correspond to two subclusters first identified by \citet{ebeling2000} and later confirmed by \citet{maughan03}. Both subclusters show similar temperatures of $\sim$5.6 keV \citep{maughan03}. A catalog of 102 confirmed cluster members has been published by \citet{demarco05}.

\subsection{CLJ1226.9+3332}

CLJ1226 has a bolometric X-ray luminosity of L$_{X}$ = 5.3 x 10$^{45}$ erg/s, an X-ray temperature of T$_{X}$ = 11.5 keV, and a total mass M = 1.4 x 10$^{15}$ M$_{\odot}$ with a virial radius of 1.66 Mpc \citep{maughan04}. \citet{ellis06} identified 45 spectroscopically-confirmed cluster members. CLJ1226 shows little dynamic activity and relaxed X-ray emission, although XMM and Chandra observations have revealed an asymmetry in the temperature distribution indicative of a merger event in the cluster \citep{maughan07}. 

\subsection{MS0451.6-0305}

MS0451 has a bolometric X-ray luminosity of L$_{X}$ = 2.01 x 10$^{45}$ erg/s and an X-ray temperature of T$_{X}$ = 10.0 keV \citep{donahue03}. The cluster has an elongated but smooth distribution of galaxies, and \citet{moran07} identified 319 spectroscopically-confirmed cluster members between 0.52 $<$ z $<$ 0.56. It has a virial radius of 2.6 Mpc \citep{moran07} and a M$_{200}$ of 1.4 x 10$^{15}$ M$_{\odot}$ \citep{donahue03}. \citet{moran07} speculate that physical processes related to the intracluster medium (ICM) which are important in the evolution of currently infalling galaxies, such as gas starvation and ram pressure stripping, begin to affect galaxies at a large radius (e.g., r $\sim$ 3.5--5.5 Mpc for merging, 0--2.5 Mpc for ram pressure stripping, 0--3.5 Mpc for starvation, and 0--5 Mpc for harassment). \citet{molnar02} identified 14 unresolved X-ray point sources in the field of MS0451, though they claim this number is within 1$\sigma$ of the number expected from a non-cluster background field.

\subsection{MS1054.4-0321}

MS1054 is an Abell class 3 cluster with a bolometric X-ray luminosity L$_{X}$ = 23.3 x 10$^{44}$ erg/s \citep{stott07} and a virial mass M$_{200}$ = 1.1 x 10$^{15} M_{\odot}$ \citep{jee052}. Substructure in the distribution of galaxies matches diffuse soft X-ray emission, and weak lensing indicates that the cluster is young, massive, and still relaxing \citep{hfk00}. \citet{tran99} found a virial radius for the cluster of 1.8 Mpc. In addition to ACS and ground-based optical and IR imaging, there have been spectroscopic surveys of this cluster using Keck \citep[e.g.,][]{vD00, tran07} to determine cluster membership of many galaxies in the ACS images. 

\subsection{SDSS1004+41}

SDSS1004 contains the quadruply-lensed background quasar SDSS J1004+4112 at z = 1.734 \citep{inada03}; there is also a faint fifth image projected through the cluster Brightest Cluster Galaxy (BCG) \citep{inada05}. \citet{williams04} modelled the cluster and calculated a virial radius of 1--1.5 Mpc and a virial mass M$_{200}$ = 4.2 x 10$^{14} M_{\odot}$. \citet{oguri10} determine that the core of SDSS1004 is highly evolved and a comparison between SDSS1004 and MACSJ1423 by \citet{limousin10} shows that the two clusters are very similar based on the distribution of their mass, light, and gas. We adapt a virial radius of 1.35 Mpc for SDSS1004 for this study, as it is the value of the virial radius of MACSJ1423 and falls in the middle of the range suggested by \citet{williams04}.

\section{Optical Variability}
\label{optvar}

Each cluster in our sample has 2--3 epochs of ACS observations in the \textit{I}-band (either F775W or F814W). Exposure times range from $\sim$4000--9000s and epochs are typically spaced one year apart (see Table 3). The high resolution of HST allows the nuclei of galaxies to be accurately targeted for small aperture photometry (r$_{aperture}$ = 0$^{\prime\prime}$.09), which allows us to minimise the contribution of light from the host galaxy when searching for nuclear variability from lower luminosity AGN which might otherwise be masked by their hosts. Eleven of the twelve clusters were also observed for one epoch in a second optical filter (F625W, F606W, or F555W). While all variability determination was done with the \textit{I}-band observations, \textit{V}-band images were used to characterise the AGN host galaxy colours when available (see Section~\ref{globalenvironment}). Table~\ref{acsobs} lists the relevant information for the ACS exposures.

\subsection{Source Detection and Photometry}
\label{MD}
\label{optcatalogs}

The calibrated data for each epoch was obtained from the HST Archive. Typically, a single epoch is comprised of 4 to 6 separate pointings which were combined using the Image Reduction and Analysis Facility (IRAF; PyRAF) tasks MULTIDRIZZLE and TWEAKSHIFTS. The field of view (FOV) for ACS is $\sim$5.2 square arcminutes, corresponding to 1.91--2.42 Mpc at redshifts of z=0.504--0.888, and the images were scaled to a resolution of 0$^{\prime\prime}$.03 per pixel. Source Extractor (SExtractor) was used to detect objects in an image created by stacking all available epochs. The catalog was then inspected to remove any spurious detections (i.e. cosmic rays, saturated stars, or objects which fell on the edge of the frame). 

\begin{figure} 
\includegraphics[width=84mm]{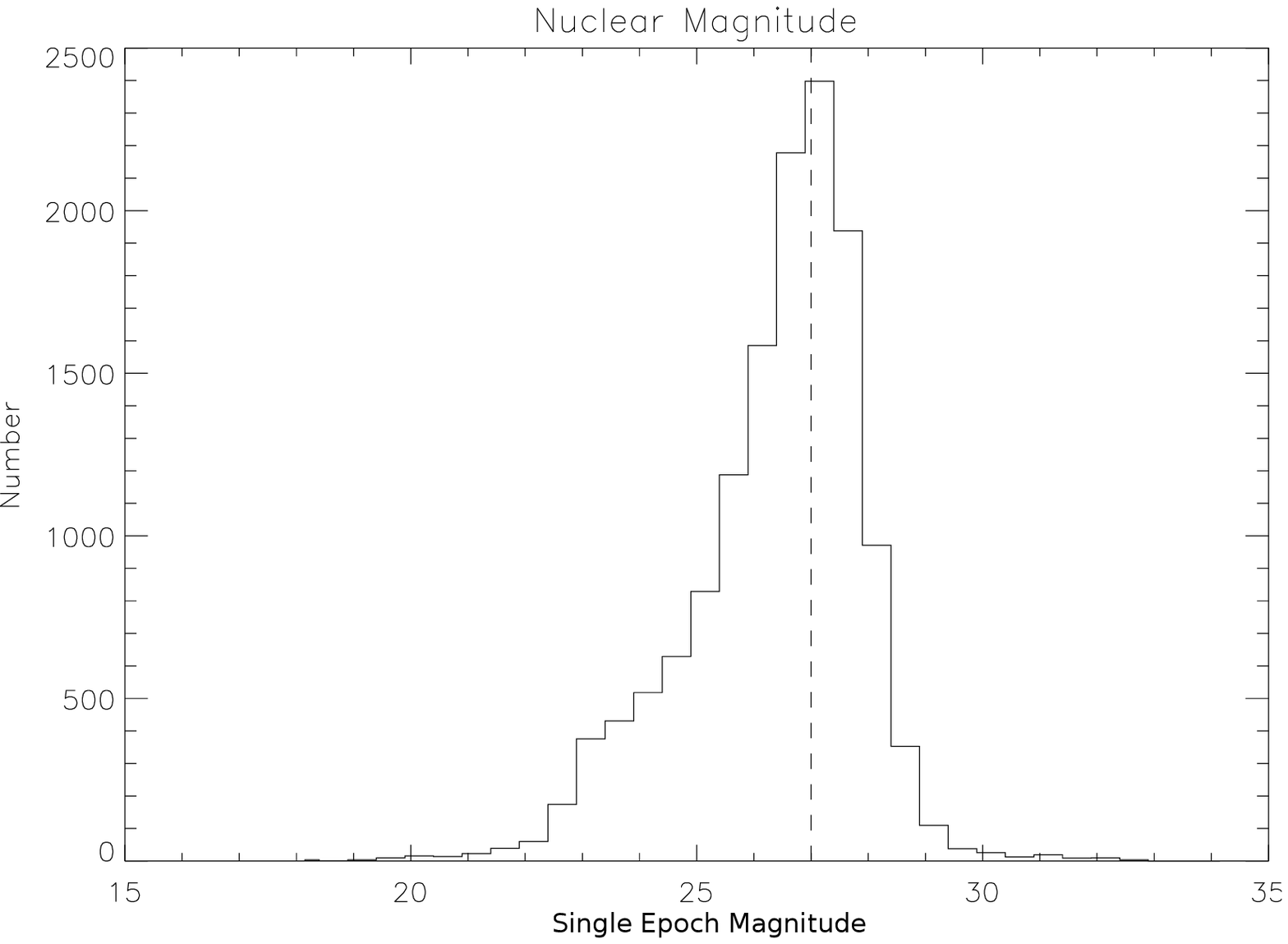}
\caption[Galaxy Nuclear Magnitude in Epoch 1]{Histogram of nuclear magnitude (\textit{I}-band) in a single epoch for all galaxies in the cluster sample. The magnitude limit is indicated by the vertical dashed line.}
\label{nucmaghist}
\end{figure}

The temporal spacing of the epochs for each cluster was approximately a year. We expect most AGN to be variable on these timescales \citep[e.g.,][]{kkc86,wm00,sesar06}. Small aperture photometry was performed on each epoch using the PHOT task in PyRAF. Based on the radial profiles of stars and compact galaxies in the fields, r = 2.5 is $\sim$2x the FWHM of an unresolved source, which is appropriate for use here as the AGN emission originates from an unresolved nuclear region. A slightly larger aperture of r = 3.0 pixels (0$^{\prime\prime}$.09) was chosen for the photometric catalog to ensure that changes in the PSF would not result in spurious detections of variables \citep[e.g.,][]{sarajedini11}. We limit our photometric survey to sources with \textit{I}$_{nuc}$ $\le$ 27. At magnitudes fainter than this limit, the number counts of sources quickly decline, indicating rapidly increasing incompleteness (Figure~\ref{nucmaghist}). We also find that the photometric noise at magnitudes fainter than this limit increases greatly, making variability detection less reliable. This flux limit corresponds to an absolute magnitude of M$_{I}$ = -15.25 at z = 0.5 and M$_{I}$ = -16.79 at z = 0.9 for the galaxy nuclei. 

\subsection{Variability Analysis}
\label{varsdet}

In order to measure optical variability, the galaxy cluster sample was divided into three groups by exposure time to maximise the total number of sources used to measure the photometric error and determine a threshold for variability. Group 1 consists of those clusters with three epochs of observations and exposure times $\sim$3000s per epoch (MACSJ0257, MACSJ0717, MACSJ0744, MACSJ0911, SDSS1004). Group 2 consists of clusters with two epochs of observations and exposure times $\sim$3300s per epoch (MACSJ1149, MACSJ1423, MACSJ2214). Group 3 consists of clusters with two epochs of observations with exposure times $\sim$2100s per epoch (CL0152, CLJ1226, MS0451, MS1054). In Section~\ref{2epochs} we describe the method of determining variability for clusters with 2 epochs of ACS observations (Groups 2 and 3), and in Section~\ref{3epochs} we describe variability determination for clusters with 3 epochs of observations (Group 1).

\subsection{Variability with Two Epochs}
\label{2epochs}

In the case of clusters with only two epochs of ACS observations, two ``fake'' epochs were constructed in the same manner as the ÒrealÓ epochs. These fake epochs were made by mixing the raw images, taking two from epoch 1 and two from epoch 2 -- resulting in two ``average'' representations of the field with similar noise and depth as the true individual epochs. These images and the resulting photometry carry through the effects of object and sky Poisson noise and readout noise. However, no intrinsic variations should exist and thus the apparent variations in photometry between them can be used to define the noise-magnitude relation upon which we base our variability threshold. Aperture photometry was performed on all sources in both ``fake'' epochs and the magnitude difference of each object was measured. We then combined the measured magnitude differences of the sources from all clusters in Group 2 and 3 to determine the photometric noise and set a threshold for variability in the real epochs for each group. This approach is the same as that used in Sarajedini et al. (2000) for the Hubble Deep Field.

\begin{figure}
\includegraphics[width=84mm]{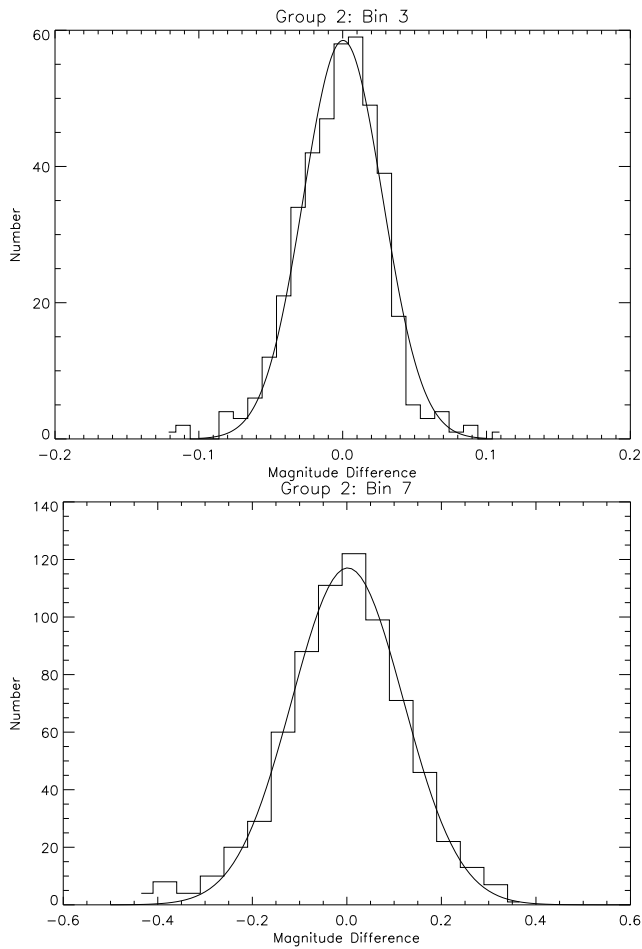}
\caption[Group 2: Galaxy Nucleus Magnitude Difference in Two Bins]{Gaussian fits to the histogram of magnitude differences in two magnitude bins for Group 2. Bin 3 (top) is the difference in magnitudes for objects between \textit{I} = 24.5--25.5 and Bin 7 (bottom) is the difference in magnitude for objects with magnitudes between \textit{I} = 27--27.5. The width of this Gaussians plotted here, $\sigma$, are used to calculate the variability threshold for sources in that magnitude bin.}
\label{examplegauss}
\end{figure}

To quantify photometric noise in our images as a function of magnitude, the magnitude differences were binned and a Gaussian function was fit to the distribution of magnitude differences in each bin. The bin sizes were chosen such that an adequate number of sources fell in each bin for an accurate Gaussian fit. The bin sizes generally spanned a full magnitude in the brighter range ($\sim$22--24) and half a magnitude among fainter objects ($\sim$24--27). Figure~\ref{examplegauss} shows an example fit to two magnitude bins for Group 2, illustrating the increased photometric error observed in faint sources compared to bright sources. 

\begin{figure}
\includegraphics[width=84mm]{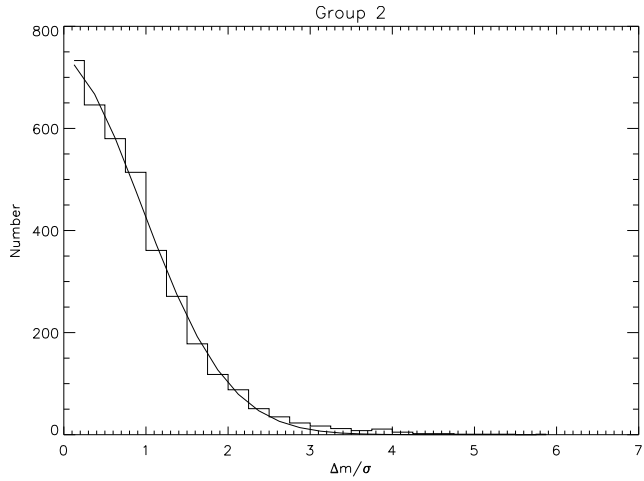}
\caption[Group 2: Determination of $\sigma^{\ast}$]{Gaussian fit to assess the photometric noise in the fake epochs in Group 2. The value of "width" for this fit is 1.06 (this is $\sigma^{\ast}$). Thus any ``3$\sigma$-confidence'' variability detection must actually display magnitude variations of 3.18 times the $\sigma$ value determined at its magnitude. The fit for Group 3 was found to have a value for $\sigma^{\ast}$ of 1.07.}
\label{examplefoldedhist}
\end{figure}

We measure the standard deviation of the Gaussian fits to the magnitude difference histograms (called $\sigma$ hereafter). We then fit a polynomial function to the $\sigma$ value to produce a smoothly changing variability threshold as a function of magnitude. The magnitude difference of each object was then divided by the value of $\sigma$ at that magnitude. For a perfect Gaussian distribution, a histogram of these values would be well fit with a Gaussian function having a width or $\sigma$ of 1 and thus any object showing a change in magnitude greater than 3$\sigma$ could be identified as varying at 3-sigma significance. We found that the width of the Gaussian fit to $\Delta$m/$\sigma$ for the photometry of all sources in the fake epochs was generally slightly larger than 1. This value, $\sigma^{\ast}$, was then adopted as a more conservative variability threshold by defining the threshold as 3.0 x $\sigma$ x $\sigma^{\ast}$, where the value of $\sigma$ varies with magnitude. Figure~\ref{examplefoldedhist} illustrates the histogram of $\Delta$m/$\sigma$ in the fake epochs for all sources in Group 2. 

\begin{figure*}
\includegraphics[width=174mm]{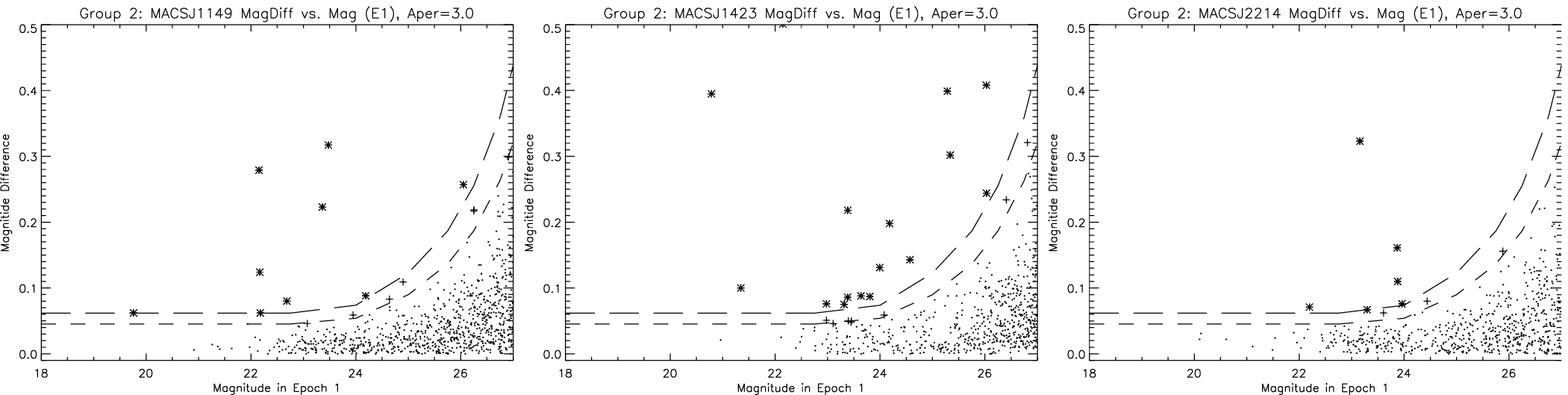}
\caption[Group 2: Optical Variability Determination]{Plots of magnitude difference vs. magnitude in epoch 1 for all clusters in Group 2. Small dots are the data; crosses are $>$3$\sigma\sigma^{\ast}$ variables and asterisks are $>$4$\sigma\sigma^{\ast}$ variables. The short and long dashed lines illustrate the 3$\sigma\sigma^{\ast}$ and 4$\sigma\sigma^{\ast}$ confidence variability thresholds.}
\label{group2}
\end{figure*}

\begin{figure*}
\includegraphics[width=174mm]{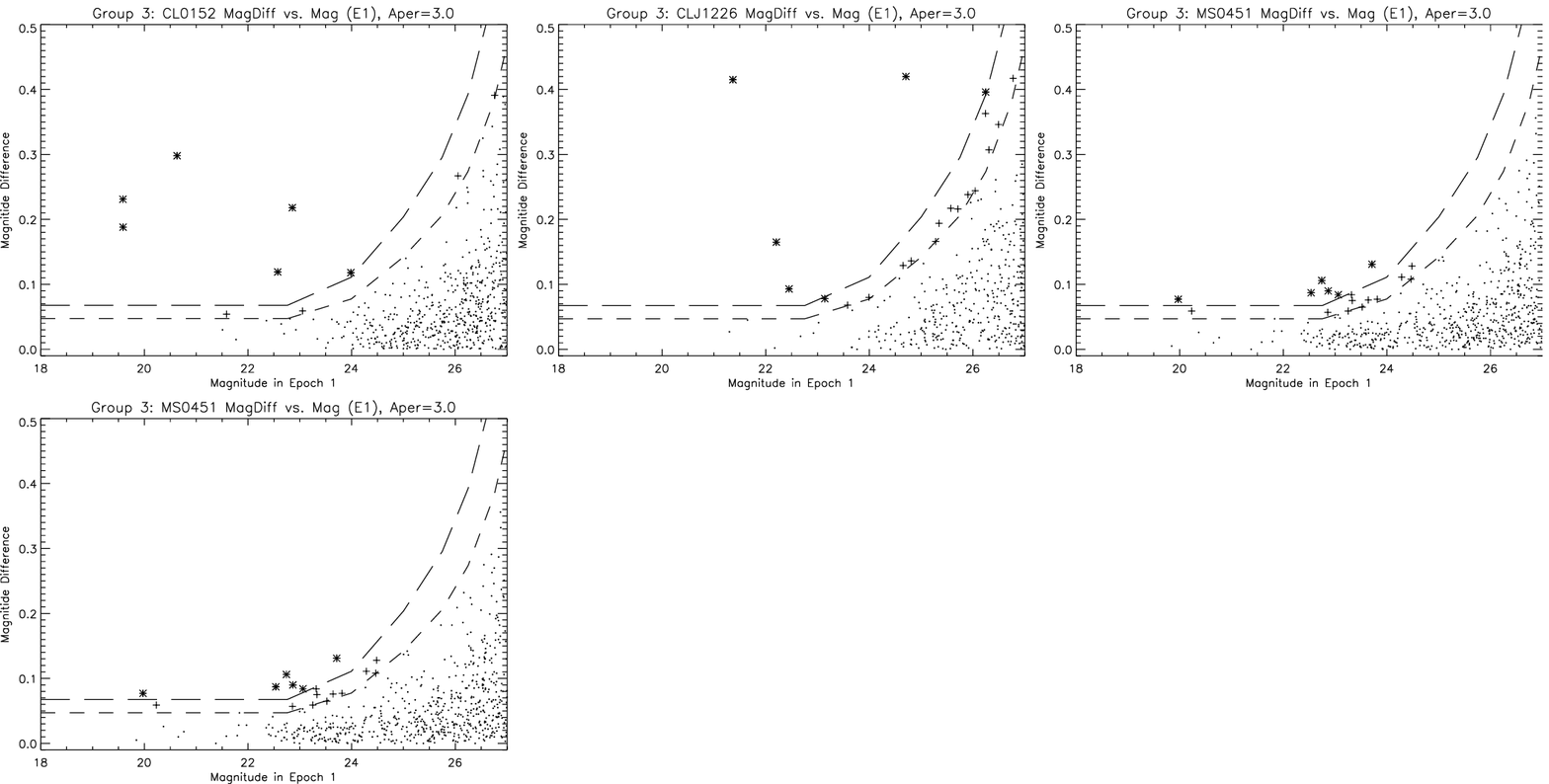}
\caption[Group 3: Optical Variability Determination]{Plots of magnitude difference vs. magnitude in epoch 1 for all clusters in Group 3. Small dots are the data; crosses are $>$3$\sigma\sigma^{\ast}$ variables and asterisks are $>$4$\sigma\sigma^{\ast}$ variables. The short and long dashed lines illustrate the 3$\sigma\sigma^{\ast}$ and 4$\sigma\sigma^{\ast}$ confidence variability thresholds.}
\label{group3}
\end{figure*}

The variability threshold determined using the fake epochs can then be applied to the true time-difference epochs, in which $\Delta mag$ (i.e. $mag1 - mag2$) was calculated and the objects sorted into bins using the same criteria as that for the fake epochs. Variables were then identified as objects with magnitude differences in the real epochs greater than the variability threshold 3$\sigma\sigma^{\ast}$. Figures~\ref{group2} and~\ref{group3} show the variability plots for all clusters in Groups 2 and 3, and Table~\ref{2evar} lists the optical variables identified in clusters with 2 epochs of HST ACS data and their properties.

\subsection{Variability with Three Epochs}
\label{3epochs}

\begin{figure*}
\includegraphics[width=174mm]{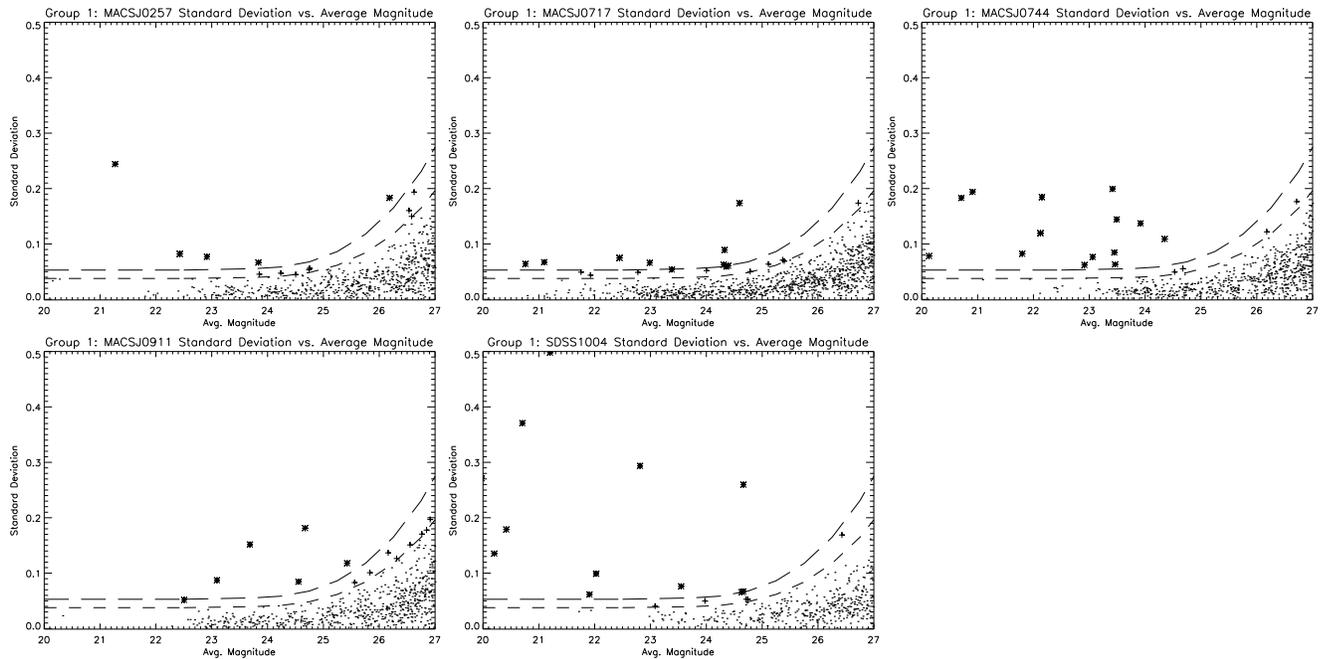}
\caption[Group 1: Optical Variability Determination]{Plots of standard deviation vs. average magnitude for all clusters with 3 epochs of ACS data (Group 1). Small dots are the data; crosses are 3$\sigma$ variables and asterisks are 4$\sigma$ variables. The short and long dashed lines illustrate the 3$\sigma$ and 4$\sigma$ variability thresholds, respectively.}
\label{group1}
\end{figure*}

In the case where three epochs of ACS data were available (Group 1), variability was determined in a different way, taking advantage of the additional epoch of imaging data. To determine the threshold of variability, the average magnitude and standard deviation were calculated for each object using the photometry from all three epochs. Objects were divided into magnitude bins and a Gaussian function was fit to the values of the standard deviation of the magnitude in each bin. A polynomial function was fit to the histogram central values (i.e., the mean value of the standard deviation as a function of magnitude). We also fit a polynomial to the Gaussian $\sigma$ values as a function of magnitude. The variability threshold was then defined as the centre of the Gaussian + 3 x $\sigma$ (where $\sigma$ is the standard deviation of the fit). Thus, an object was defined as variable if the measurement of its magnitude showed a standard deviation above this threshold within its magnitude bin. This approach is similar to that used in \citet{ks07} and in \citet{sarajedini11} for their 5-epoch variability analysis of the GOODS fields. Figure~\ref{group1} shows the variability plots for all clusters in Group 1, and Table~\ref{3evar} lists the optical variables identified in clusters with 3 epochs of HST ACS data and their properties.

\subsection{Summary: Optical Variables in Clusters}
\label{varsummary}

We identified two known supernovae among our selected variables from the catalog of \citet{sharon2010} and removed these, as our aim is to study the AGN population. We find 178 AGN candidates based on the detection of optical variability in our sample of 12 clusters. Ninety (51\%) of these variables have greater than 4$\sigma$ significance. We detect an average of 15 variables per cluster, with a range of 8--24 variables per cluster. We find that 1.1\% (178/15,849) of all galaxies in our cluster images are variable down to \textit{I}$_{nuc}$ = 27. 

The greatest sources of incompleteness in our variability sample is due to undersampling of the AGN lightcurve with only 2 or 3 epochs of data. Previous variability surveys which have imaged the same field several times over the course of several years \citep[e.g.,][]{trevese94,hawkins02} have found that virtually all AGN vary over such a temporal baseline when several epochs of data are analysed. In our survey, we sample only 2 or 3 points on the AGN lightcurve over $\sim$1 year, which should result in some incompleteness in our variability survey. In a study of the Hubble Deep Field using 2 epochs of data separated by a few years, \citet{sarajedini03} estimate that $\sim$75\% of variable AGN would be detected. We expect this same level of completeness in our survey, as we have a similar number of epochs and similar temporal coverage.

\begin{figure} 
\includegraphics[width=84mm]{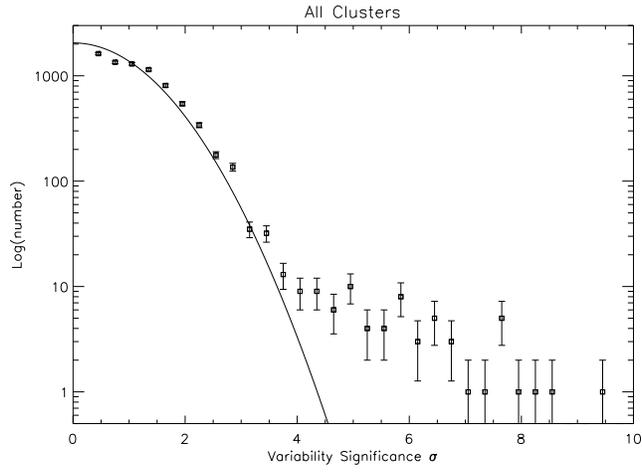}
\caption[Distribution of Variability Significance]{Distribution of $\sigma$ for all galaxies in our clusters. The x-axis shows variability significance for all galaxies, normalised to remove $\sigma^{\ast}$ in Groups 2 and 3. The y-axis shows the logarithm of the number of galaxies in bins of 0.3. Error bars represent the Poisson statistical errors in each bin. The solid line is a Gaussian fit to the data within 3$\sigma$.}
\label{histsigma}
\end{figure}

False positives or spurious variable sources are also an issue in variability selected samples. To estimate the number of expected spurious sources in our sample, we examine the distribution of variability significance values for all galaxy nuclei in 12 cluster images. Figure~\ref{histsigma} shows a histogram of variability significance for all nuclei. The data are well fit by a Gaussian distribution out to a variability significance of 3 and this part of the histogram represents the large number of non-varying galaxy nuclei in our survey. The significant variables, however, appear as a separate population extending to higher significance values. In a normal distribution, we would expect that $\sim$48 sources from among our total survey would have significance values greater than 3. Based on these statistics, we estimate that $\sim$30--40 sources (17--22\% of our variables) could be spurious detections. 

\section{X-ray Data Reduction and Point Source Detection}
\label{xdet}

AGN are known to be luminous X-ray sources and can be identified from deep X-ray surveys. Although many previous X-ray surveys for point sources in galaxy clusters have been hindered by the emission of hot intracluster gas, the current generation of X-ray observatories offer the sensitivity, resolution, and positional accuracy needed to detect point sources even in the cores of clusters, and wavelet detection techniques \citep[e.g.,][]{freeman02} allow for accurate separation of point source emission from the diffuse background. Thus, the Chandra X-ray Observatory and its supported software package, CIAO, allow us to search for X-ray point sources in the cores of the galaxy clusters in our survey sample.

\subsection{Data Reduction}

Archival X-ray observations taken with the Chandra X-ray Observatory were obtained from the Chandra Data Archive. Table~\ref{xobs} lists the observation IDs and effective exposure times for each cluster. Data were prepared for analysis using CIAO 4.1.1 and following the steps in the ACIS Data Preparation in the Analysis Guide including afterglow removal, flagging of bad pixels, and removal of background flare events. The observations were filtered by energy into full (0.5--8 keV), hard (2--8 keV), and soft (0.5--2 keV) bands and separate epochs were reprojected and merged into a single observation to obtain the deepest X-ray data possible, resulting in the final effective exposure times listed in Table~\ref{xobs}. The data were trimmed to the size of the ACS field of view in order to run source detection on only the area covered by the ACS optical data available for comparison with variability-selected AGN. 

\subsection{Source Detection}

We detected X-ray point sources using the CIAO tool {\fontfamily{pcr}\selectfont wavdetect} in the full (0.5--8 keV) band of the merged X-ray images. Source fluxes in the full, hard, and soft bands were calculated using the CIAO tool {\fontfamily{pcr}\selectfont eff2evt} in CIAO version 4.3, which calculates the flux in erg/cm$^{2}$/s for a specified location on the chip, taking into account the quantum efficiency and effective area.

The flux was measured within a circular aperture created around each source with a radius equivalent to the source radius determined by {\fontfamily{pcr}\selectfont wavdetect}. The background flux at the position of each source was measured using an annulus 2--5 times the size of the source radius, and subsequently subtracted from the source flux. Fluxes were calculated in each band (full, hard, soft) in each individual observation ID; the results were summed to obtain the total source flux, with weighting based on the exposure time of each observation. 

\begin{figure} 
\includegraphics[width=84mm]{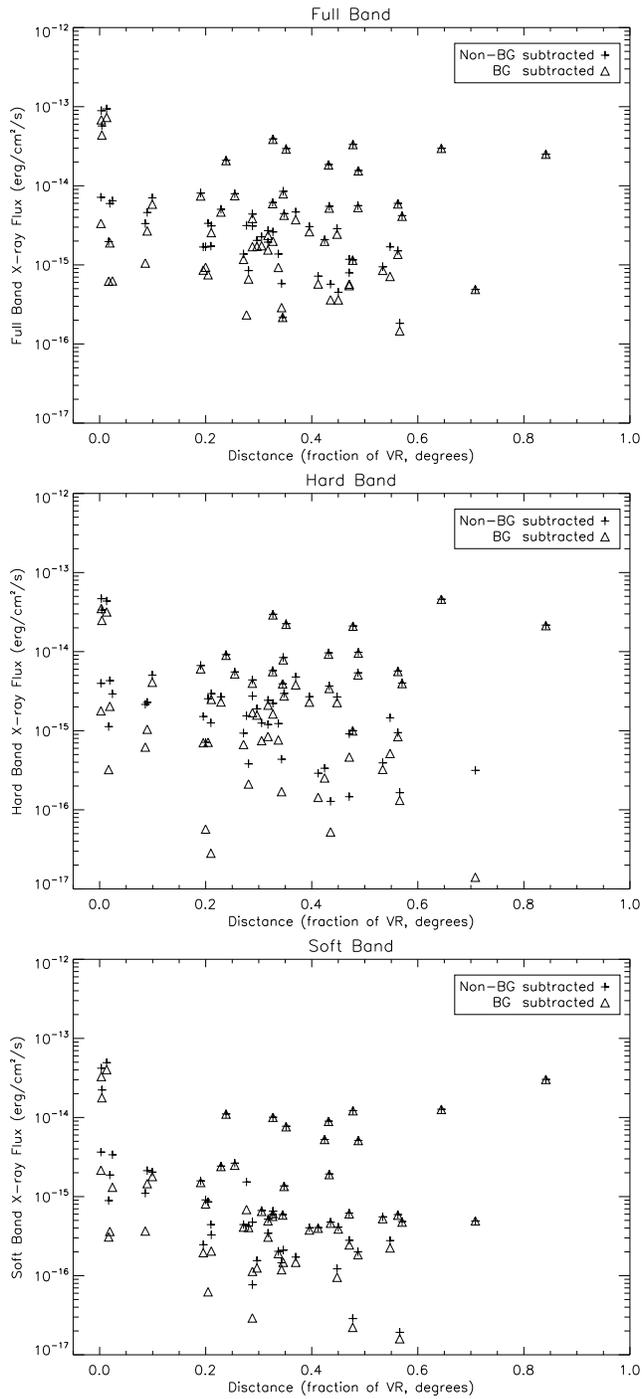}
\caption[X-ray Flux vs. Cluster Radius]{X-ray flux vs. distance from the center of the cluster (as a fraction of the virial radius in degrees) in the full (top), hard (middle), and soft (bottom) bands. Plus symbols are the measured fluxes, while triangles represent the background-subtracted flux (used throughout this work).}
\label{fluxvsdist}
\end{figure}

For the purpose of comparison with other work, we calculated rest-frame X-ray fluxes assuming a 1D power law plus Galactic absorption model and assumed all X-ray sources to be at the cluster redshift. We used the foreground Galactic extinction toward each cluster from \citet{DL90} and assumed a $\Gamma = 1.7$ power law typical of X-ray-selected AGN \citep[e.g.,][]{martini06}, where $\Gamma$ is the slope of the power law photon flux density $N_{E} \propto E^{-\Gamma}$. K-corrections to the observed fluxes were calculated using the {\fontfamily{pcr}\selectfont calc$\_$kcorr} function of the Sherpa package in CIAO 4.3. Figure~\ref{fluxvsdist} shows the X-ray flux vs. distance from the center of the cluster (defined to be at the position of the Brightest Cluster Galaxy, BCG) as a fraction of the virial radius. While the figure shows both the measured source flux (plus symbols) and the background-subtracted source flux (triangles), only the background-subtracted flux is considered throughout this work. It can be seen that the background subtraction is significant in the inner parts of the clusters, where X-ray emission from hot intracluster gas is significant; the background subtraction is small in the outer parts of the clusters, where the contribution from this gas is minimal. This figure demonstrates our ability to detect X-ray sources down to approximately the same completeness limit in the inner regions of the clusters, where extended cluster X-ray emission is present.

\subsection{Summary: X-ray-Detected Sources in Clusters}

We find 74 X-ray point sources in our 12 clusters. This is an average of 6 X-ray point sources per cluster with a range of 1--12 per cluster over our entire sample. X-ray point sources in each cluster and their properties are listed in Table~\ref{xpsprop}. 

\begin{figure} 
\includegraphics[width=84mm]{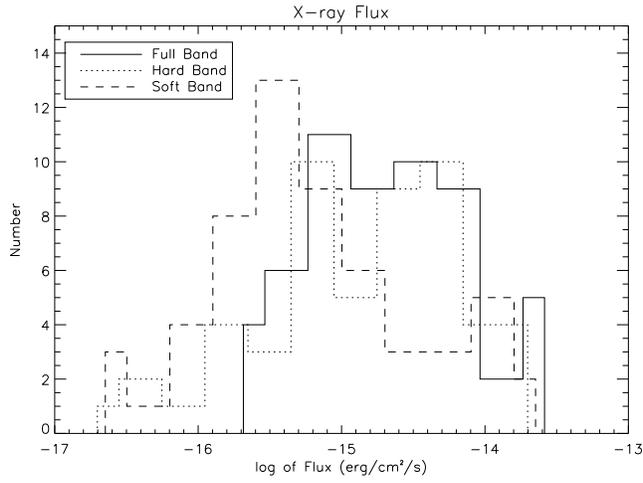}
\caption[Distribution of X-ray Flux]{Histogram of the log of X-ray flux (erg/cm$^{2}$/s) in the full, hard, and soft bands.}
\label{xfluxhist}
\end{figure}

Figure~\ref{xfluxhist} shows a histogram of the X-ray flux in our three bands. We are able to detect sources in the full band to a flux of $\sim$7 x 10$^{-16}$ erg/cm$^{2}$/s. This flux corresponds to an X-ray luminosity of $\sim$6 x 10$^{40}$ erg/s at a redshift of 0.5 and $\sim$3 x 10$^{42}$ erg/s at a redshift of 0.9. Our survey is thus sensitive to most AGN across the redshift range of our cluster sample. Table~\ref{xrayfit} lists the parameters used to calculate X-ray luminosities for our point source sample.

We calculate X-ray hardness ratios for objects detected in both the soft (0.5--2 keV), and hard (2--8 keV) bands. Following the same procedure used for the merged full band images, {\fontfamily{pcr}\selectfont wavdetect} was used to analyse the merged soft and hard band images. For objects detected in both bands, we calculate a hardness ratio HR \citep[e.g.,][]{martini02}:

\begin{equation}\label{} HR = \frac{F_{X}(2-8keV)}{F_{X}(0.5-2keV)}\end{equation}

\begin{figure} 
\includegraphics[width=84mm]{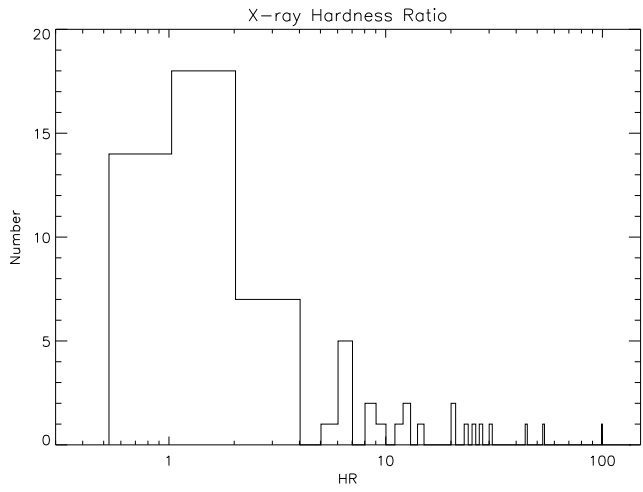}
\caption[Distribution of X-ray Hardness Ratio]{Histogram showing the distribution of X-ray hardness ratios for our sample of X-ray-detected AGN.}
\label{hr}
\end{figure}

\noindent
Figure~\ref{hr} shows the distribution of hardness ratios for the 69 X-ray point sources which are detected in both the hard and soft bands. We find that most sources have HR values on the soft end of the distribution, with 46 sources (67\%) having hardness ratios $\le$ 5.

\section{Infrared Image Analysis and Photometry}
\label{irdet}

If dust is present around the AGN, X-ray, UV, and optical emission can be absorbed and re-radiated as infrared light. Galaxies whose emission is dominated by an obscured AGN tend to have mid-IR SEDs which follow a power law ($f_{\nu} \propto \nu^{\alpha}$, where $\alpha$ is the spectral index). \citet{donley08} find that a power law fit to mid-IR photometry produces a more reliable AGN sample compared to IR colour selected samples or mid-IR excess selection. AGN identified via a power-law fit to the SED in the mid-IR produces a catalog of objects that may not be identified via optical or X-ray emission due to obscuration. 

\subsection{Data Reduction}

Archival Spitzer observations taken with the IRAC (Infrared Array Camera) instrument were available for seven clusters (indicated in Table~\ref{clustersgen}) with exposure times ranging from $\sim$900--6000s. All observations were processed using the IRAC pipeline (version S18.0 or later), and thus corrected BCD (basic calibrated data) files known as cBCD files were available from the Spitzer Science Center. These images were corrected for artefacts such as muxbleed (residual signal in multiplexers), column pulldown (bias shift in columns containing bright sources), banding (bias shift in rows containing bright sources), and first-frame correction (temporal dependence of the bias of each array). 

The data were obtained using the Spitzer Pride program Leopard and mosaicked into images for photometry using the MOPEX (MOsaicker and Point source EXtractor) tool which processes calibrated images into a science grade mosaic for point source extraction and photometry. Point source extraction can also be performed with MOPEX, but was in this case performed with Source Extractor in order to incorporate catalogs derived from the optical and X-ray observations. 

\subsection{Source Identification and Photometry}

Source Extractor was used to identify sources and measure the magnitude in each IRAC band, using the 3.6$\mu$m band as a reference catalog. Photometry was performed with apertures of radius 3$^{\prime\prime}$.6 (6 pixels, or 3 ``native'' IRAC pixels) and a background annulus from 3$^{\prime\prime}$.6 -- 8$^{\prime\prime}$.4 (8 pixels, 3--7 ``native'' IRAC pixels), following the suggestions in the IRAC instrument handbook.\footnote{http://ssc.spitzer.caltech.edu/irac/iracinstrumenthandbook/} The photometric catalog was further refined to include only objects detected in all four IRAC wavebands using the SPHEREMATCH routine in IDL. The flux of each source was calculated by converting the flux from SExtractor (in MJy/sr) to microJy/pix and applying a specified aperture correction from Table 4.7 in version 1.0 (February 2010) of the Spitzer IRAC Instrument Handbook.\footnote{http://ssc.spitzer.caltech.edu/irac/iracinstrumenthandbook/home/} The zeropoints to convert flux into Vega magnitudes were obtained from Appendix B (Performing Photometry on IRAC Images) of the Instrument Handbook.

\begin{figure} 
\includegraphics[width=84mm]{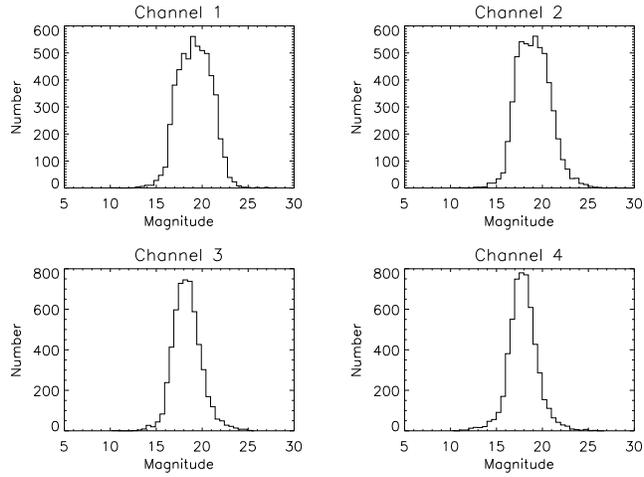}
\caption[Distribution of IR Magnitude]{Histograms of IR magnitudes in each of the four Spitzer IRAC channels.}
\label{IRmaghist}
\end{figure}

Figure~\ref{IRmaghist} shows a histogram of object magnitudes in each of the IRAC channels. A magnitude of 18 in Channel 4 (8 $\mu$m) corresponds to an IR luminosity of L$_{IR, 8 \mu m}$ = 1.4 x 10$^{42}$ erg/s at a z = 0.5 and L$_{IR, 8 \mu m}$ = 6.0 x 10$^{42}$ erg/s at z = 0.9. The flux values over all four IRAC channels was fit with a power-law of the form 

\begin{equation}\label{irfluxeq} f_{\nu} \propto \nu^{\alpha}\end{equation}
\noindent
and minimised $\chi^{2}$, selecting galaxies that are well fit (0.5 $<$ $\chi_{reduced}^{2}$ $<$ 1.1) with a spectral index $\alpha$ $<$ -0.5 within 1$\sigma$ of their errors, following the selection criteria of \citet{ah06} for the identification of AGN candidates. This follows from the work of \citet{ivezic02}, which shows that the spectral indices of optically-selected quasars in the Sloan Digital Sky Survey are in the range of -0.5 $<$ $\alpha$ $<$ -2. \citet{ah06} also find that the slope of the power law fit relates to the AGN type, where steeper (i.e., more negative) values represent NLAGNs and shallower power law SEDs are classified as BLAGNs. 

\begin{figure}
\includegraphics[width=84mm]{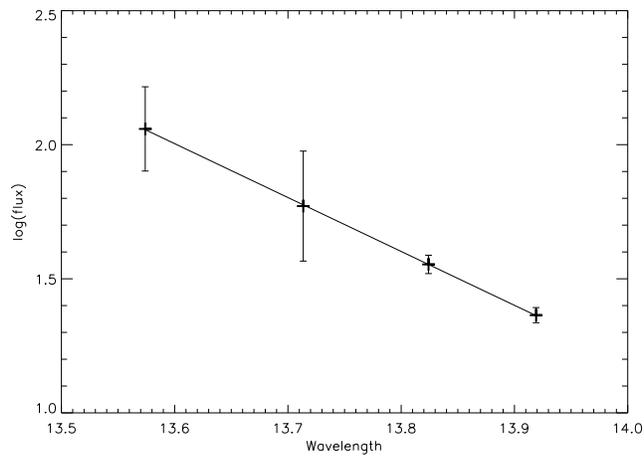}
\caption[Example Power Law Fit to a Mid-IR SED]{Example of a galaxy which shows an SED over the four IRAC channels which fit a power law model with a spectral index $\alpha$ between -0.5 and -2.}
\label{explaw}
\end{figure}

Figure~\ref{explaw} is an example of one of our mid-IR power law-selected AGN. Table~\ref{plawprop} lists the IRAC power law AGN candidates identified in the seven clusters in our study with Spitzer observations along with their properties. 

\subsection{Summary: IR Power-Law Sources in Clusters}

We find 64 sources in our seven clusters with Spitzer observations displaying mid-IR power-law SEDs. This results in an average of 9 IR power-law sources per cluster with a range of 2--18 per cluster. Of our 64 sources, 20 (31\%) have slopes steeper than $\alpha$ = -0.9 and are similar to the NLAGN SEDs of \citet{ah06}. The remaining 44 (69\%) have shallower slopes and more closely match the BLAGN SED template. We explore the mid-IR colours of these and AGN selected via optical variability and X-ray emission in the following section.

\section{Multi-wavelength Analysis of AGN Candidates}
\label{overlaps}

The catalogs of X-ray and mid-IR AGN candidates were matched against the optical catalog using the IDL routine SPHEREMATCH with a 2$^{\prime\prime}$ (67 pixels) maximum match radius. While there was generally good alignment between the X-ray and IR data, linear offsets were quantified and applied to the catalogs where needed. 

Table~\ref{overlap} lists the IR- and X-ray-selected AGN with optical counterparts in our cluster photometric survey, as well as the number of AGN candidates detected in more than one of the three catalogs. We find that 2/48 (4\%) of mid-IR power law sources are detected as optical variables and one of these is variable at $>$4$\sigma$ confidence. The overlap is greater among the X-ray population, where we find that 12/50 (24\%) X-ray point sources are optical variables and two-thirds of these are variable at $>$4$\sigma$ confidence. 

We compare these results to those of \citet{sarajedini11} and \citet{ks07}, which examine the properties of AGN candidates identified using similar techniques in the GOODS fields. Both of these studies found that optical variability was observed in about one-quarter of X-ray sources and about half of all mid-IR power-law sources. These percentages do not change considerably when we consider only GOODS field X-ray sources bright enough to be detected in the shallower Chandra observations available for our clusters (F$_{X}$ $\gtrsim$ 2 x 10$^{15}$ erg/s/cm$^{2}$ in the full band). Among the X-ray sources detected in our cluster survey, we find that 24\% exhibit optical variability, similar to that found in the field population. However, a significantly lower fraction of mid-IR power law sources (4\%) are identified as variable in clusters compared to the field.

This inconsistency may be at least partially explained by the different type and temporal sampling of optical data used to determine optical variability in the two surveys. In the GOODS, variability was measured using \textit{V}-band (F606W) photometric data with 5 epochs covering a 6-month time span. In the cluster analysis, \textit{I}-band (F775W and F814W) photometry was used since multi-epoch imaging was only available at this wavelength. In addition, variability was determined from just 2 or 3 epochs, though generally over a longer time baseline of 1 or 2 years. We expect our variability analysis to be somewhat less sensitive to varying nuclei at this longer wavelength, since it has been shown that optical variability amplitudes increase with decreasing wavelength \citep{vb04}. Based on a comparison of \citet{sarajedini00} and \citet{sarajedini03}, we estimate that about $\sim$50\% of \textit{V}-band-detected variables would be identified as variables in our \textit{I}-band survey. Thus, the lower percentage of mid-IR sources identified as variable in our survey may be partially explained by less sensitivity to weakly-varying nuclei in the \textit{I}-band. Indeed, this is seen in Figure 2 of \citet{ks07}, which shows that the variability significance of mid-IR sources is generally quite low, indicating low-level amplitudes of variability. The X-ray sources do not show the same level of sensitivity, which may be due to the fact that they generally display overall greater variability amplitudes (as demonstrated in the large number of $>$4$\sigma$ confidence variables among this population and also shown in Figure 2 of \citet{ks07}).


We find that $\sim$7\% of variables are also identified as AGN candidates through X-ray emission or mid-IR power law fits. This is less than the fraction of field galaxy variables that are also X-ray/mid-IR sources (36\% of GOODS variables after correcting for the X-ray flux level of the cluster survey). This difference may also be due to the lower sensitivity of the \textit{I}-band variability survey in which the light may be more dominated by stars and incompleteness due to the small number of epochs as discussed in Section~\ref{varsummary}.

When comparing our IR and X-ray catalogs, we find that 6/64 (9\%) of all IR power law sources are detected as X-ray point sources; when considering only objects with optical counterparts this percentage remains roughly the same (4/48, 8\%). Approximately the same percentages are found among X-ray point sources that are also IR power law sources ($\sim$8\%). The overlap here is less than that found among deeper IR power law source surveys conducted in the GOODS fields \citep{ah06,donley07}, where about half of all galaxies displaying power law behaviour across the IRAC channels are also X-ray-detected. \citet{atlee11} recently report on a study of low-z clusters and find that just $\sim$20\% of AGN detected via IR SEDs and/or X-ray emission are identified using both techniques. They hypothesise that the lack of X-ray emission from most of the IR AGNs is due to the larger column densities of cold gas in the hosts of these galaxies, having found that IR AGN hosts have larger specific star formation rates than the hosts of X-ray AGN. A moderately large column density of cold gas could suppress the X-ray emission from the IR AGNs, making them undetectable in our survey. Additionally, we note that the sensitivity of cluster X-ray surveys may be impeded by the underlying X-ray emitting cluster gas, which could result in some differences in the number of IR AGN detected in X-rays in clusters than in the field.

\begin{figure} 
\includegraphics[width=84mm]{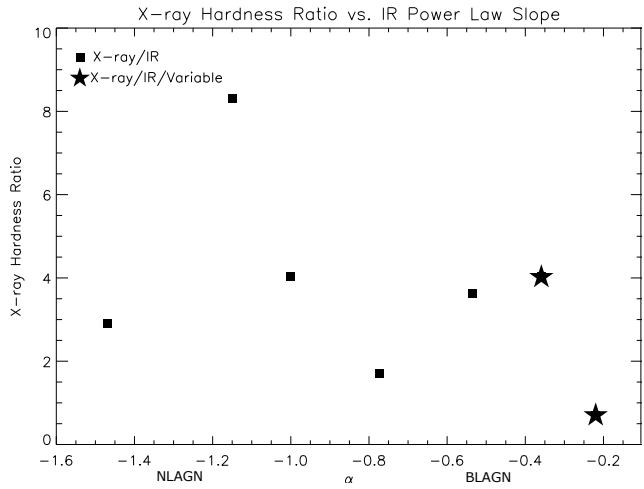}
\caption[X-ray Hardness Ratio vs. IR Power Law Slope]{X-ray hardness ratio vs. IR power law slope for AGN detected in both IR and X-rays (squares). Stars denote the 2 objects which also show optical variability.}
\label{alphahr}
\end{figure}

Figure~\ref{alphahr} plots X-ray hardness ratio vs. IR power law slope for the 7 AGN we detect via both methods. Of the AGN we detect in the mid-IR and X-rays, 4/7 have power law indices indicative of BLAGN ($\alpha$ $>$ -0.9) and 6/7 have softer X-ray hardness ratios (HR $<$ 5). The object with the hardest X-ray flux is also classified with a NLAGN-like mid-IR SED. Both of the objects also detected via optical variability have shallow, BLAGN-like SEDs and are primarily soft X-ray sources. We find no clear correlation between X-ray hardness and mid-IR SED classification, which is consistent with previous studies \citep[e.g.,][]{barmby06,ks07}. Variations in the gas-to-dust ratio or a range of intrinsic AGN properties may explain the reason for the poor correlation between measures of X-ray and IR obscuration, despite the expectation that obscuration should in principle have an effect within both wavelength regimes \citep{barmby06}. 

\begin{figure} 
\includegraphics[width=84mm]{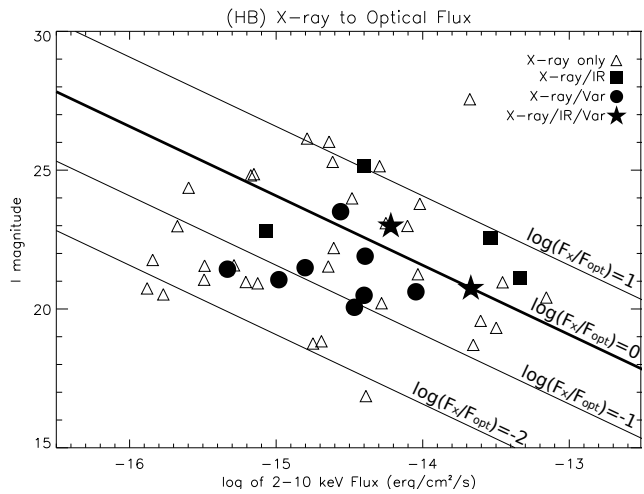}
\caption[Optical to X-ray Flux Ratio]{Optical to X-ray flux of X-ray point sources with identifiable optical counterparts; the lines indicate constant flux ratios of $log$(F$_{X}$/F$_{opt}$) = 1, 0, -1, and -2. Objects toward the top of this plot are more optically obscured, and objects toward the bottom are optically bright and X-ray weak.}
\label{xoptflux}
\end{figure}

Next, we examine the X-ray to optical flux ratios of our X-ray-detected AGN. From our catalog of 74 X-ray point sources, 50 have optical counterparts for which the \textit{I}-band magnitude was measured. We plot the X-ray to optical flux ratio of these sources in Figure~\ref{xoptflux} with lines of constant $log$(F$_{X}$/F$_{opt}$). AGN are known to occupy $log$(F$_{X}$/F$_{opt}$) = $\pm$1 \citep[e.g.,][]{comastri02,rigby06,georgakakis04}. Most (70\%) of our X-ray sources lie within this region of the diagram, consistent with the presence of an AGN. Both galaxies we detect via all three AGN detection methods and all galaxies detected in both X-rays and via a mid-IR power law SED are within this range. Of the eight objects detected via both variability and X-rays, 5 (63\%) also occupy this region. 

High ratios of X-ray to optical flux can indicate the presence of an obscured AGN, even without optical evidence for nuclear activity \citep[e.g.,][]{fiore00,hornschemeier01,giacconi01,barger01}. Low values indicate X-ray-weak sources that may be optically bright. Objects with $log$(F$_{X}$/F$_{opt}$) $\lesssim$ -2 are consistent with values of F$_{X}$/F$_{opt}$ found in star-forming galaxies \citep[e.g.,][]{moran99,alexander02,bauer02,georgakakis03}. \citet{georgakakis04} finds that objects with F$_{X}$/F$_{opt}$ below -2 have optical spectra dominated by the host galaxy, though the presence of a low-luminosity AGN (LLAGN) cannot be ruled out. X-ray bright optically-normal galaxies (XBONGs) in which the light from the host galaxy dominates the optical spectrum tend to occupy the region -2 $\lesssim$ $log$(F$_{X}$/F$_{opt}$) $\lesssim$ -1 \citep{comastri02,rigby06}. While most of our sources with $log$(F$_{X}$/F$_{opt}$) $<$ -1 are only detected in X-rays, three AGN candidates are also detected as optical variables, thus increasing the likelihood that they host an AGN and the X-ray emission we detect is due to an accreting supermassive black hole rather than processes related to stars and star formation. Such sources are also found in field AGN surveys \citep{trevese08,sarajedini11}. 

\begin{figure} 
\includegraphics[width=84mm]{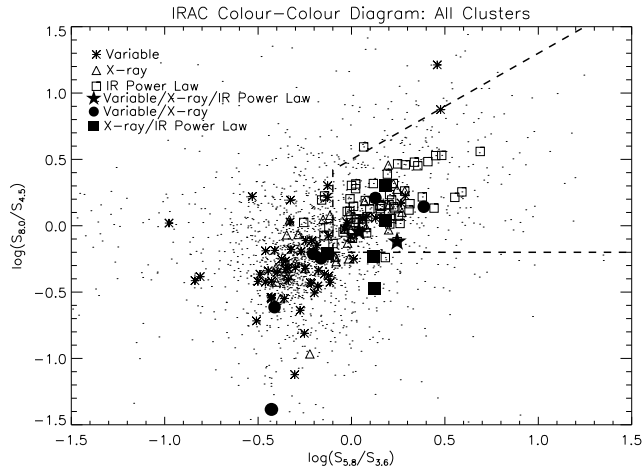}
\caption[Mid-IR Colours of Galaxies]{IRAC colour-colour plot for galaxies in all 7 clusters with Spitzer observations. The region within the dashed line is that used to select AGNs via the criteria of \citet{lacy07}. Small points are galaxies with IR emission in all four IRAC channels, and different symbols denote the AGN detected via different techniques (see legend).}
\label{IRcolors}
\end{figure}

Finally, we consider the infrared properties of our sample of AGN to determine whether the mid-IR SED is dominated by emission from the AGN or the host galaxy light. \citet{lacy04} examine the mid-IR colours of known quasars identified in the Sloan Digital Sky Survey (SDSS) to define a region in colour space for galaxies dominated by an AGN. In Figure~\ref{IRcolors} we plot our cluster galaxy colours and quantify the numbers of galaxies found in the Lacy wedge in Table~\ref{lacywedge}. We find that 89\% of IR power law sources are in the Lacy wedge, along with 16\% of optical variables and 44\% of X-ray point sources. Conversely, 11\% of IR power law sources, 84\% of optical variables, and 56\% of X-ray point sources are not in the Lacy wedge, and thus we can assume their mid-IR light is dominated by the host galaxy and star formation rather than the AGN. \citet{atlee11} find that galaxies in the AGN region of mid-IR colour-colour diagrams must have more than 50\% of their mid-IR light contributed by the AGN component. About one quarter of the X-ray and mid-IR selected AGN in their study have colours within the AGN wedge, which is in general agreement with our survey results. This is also consistent with the results of \citet{hickox09} for field galaxies, in which they find that 32\% of X-ray sources have IR colours within the Stern wedge \citep{stern05}. Thus, we find no significant difference between the percentage of X-ray objects with IR colours indicative of AGN emission between the field and cluster population.

In summary, we find that optical variability, X-ray emission, and mid-IR power law SEDs produce a robust catalog of AGN candidates in clusters, with varying degrees of overlap among the samples. Considering AGN detected in all three techniques, we find an average of 25 AGN candidates per cluster with a range of 12--49 per cluster over the survey.


\section{Determination of Cluster Galaxies}
\label{globalenvironment}

A primary goal of this work is to determine how the cluster environment affects the onset and fuelling of accretion onto a galaxy's supermassive black hole. In this section we discuss the methods by which we estimate cluster membership for galaxies in each cluster field and compare the percentage of cluster AGN with that found among field studies, as well as examine the effect of cluster physical properties on the AGN fraction.

\subsection{Spectroscopic Catalogs}

Spectroscopic catalogs of varying completeness are available from the literature for the clusters in our sample. Cluster members were determined using the published redshifts available for each cluster. In the case of the MACS clusters, this range was calculated using the measured velocity dispersion of the cluster galaxies \citep{barrett06} with an average of $\sim$40 galaxies per cluster. Table~\ref{zsandranges} lists the number of sources with published redshifts that fall within the ACS field-of-view, as well as the redshift range spanned by cluster members from the literature. We note that this redshift range is simply the lowest to highest redshift of confirmed cluster members given in the literature and does not fully represent the cluster redshift range.

These catalogs were matched with our photometric catalogs using the IDL routine SPHEREMATCH with a 2$^{\prime\prime}$ (67 pixels) maximum match radius. In some cases a linear offset between the world coordinate systems (WCS) of the ACS images and the spectroscopic catalog was applied. Table~\ref{agnzs} lists the AGN candidates in each ACS image having a measured spectroscopic redshift, confirming whether it is a cluster member or field galaxy. A dash indicates that data is not available for this cluster.

\subsection{Field Contamination}
\label{fieldcont}

Each of our cluster fields contains a combination of the galaxy cluster population together with foreground and background galaxies. In order to supplement the existing spectroscopic cluster membership information and better estimate contamination from the field galaxy population, we estimate the number density of the field population in each of our clusters. To do this, we obtained archival HST ACS images of the GOODS-North and GOODS-South fields taken in the F775W filter. Three ACS images near the center of the GOODS-N and GOODS-S fields were selected from the archive with exposure times similar to the depth of the cluster data: Field 1 (total exposure time 5000s), Field 2 (total exposure time 7028s), and Field 3 (total exposure time 8350s). The data were reduced using the same technique as the cluster ACS images. Images were drizzled to the same resolution (0$^{\prime\prime}$.03/pix). 

Source extraction and photometry using a Kron flexible elliptical aperture of all galaxies in each tile was performed with Source Extractor. The resulting galaxy photometry catalogs provide an estimate of the mean number of galaxies/arcmin$^{2}$ observed at the depth of our cluster survey: \\

\noindent 
Field 1 (5000s): 15.6 galaxies/arcmin$^{2}$ \\
Field 2 (7028s): 16.6 galaxies/arcmin$^{2}$ \\
Field 3 (8350s): 18.4 galaxies/arcmin$^{2}$ \\

For each cluster, we now have an estimate of the density of the field galaxy population. We use this to estimate cluster membership probabilities for sources without spectroscopic information as described in the following section.

\subsection{Cluster Radial Profiles and Membership Probability}
\label{radialprob}

\begin{figure} 
\includegraphics[width=84mm]{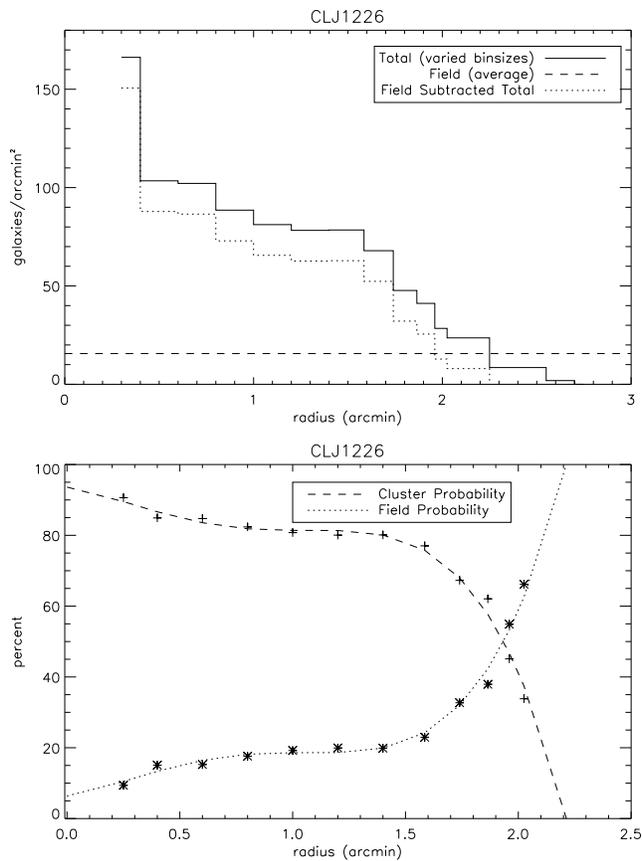}
\caption[Radial Distribution of Galaxies in CLJ1226]{{\it Top Panel}: Radial plot of CLJ1226 showing galaxies/arcmin$^{2}$ vs. radial distance from the center in arcminutes. The solid histogram is the total number of objects in each radial bin and the dotted histogram is the field-subtracted radial profile of the cluster. The average number of galaxies/arcmin$^{2}$ is indicated with a horizontal dashed line. {\it Bottom Panel}: Radial plot of CLJ1226 showing the probability that a galaxy at this radius resides in the cluster (plus symbols) and the field (asterisks) as a function of distance from the center in arcminutes. The dashed and dotted lines are polynomial fits to the cluster and field probabilities.}
\label{exampleradprof}
\end{figure}

Assuming that the center of each cluster (defined as the Brightest Cluster Galaxy) contains the highest concentration of cluster galaxies with a decreasing contribution from the galaxy cluster and increasing contribution from the field with increasing distance from the center, we calculate the number of galaxies/arcmin$^{2}$ vs. radius in bins of varying size from 0.3--0.01 arcmin starting from the center of each cluster. We then assume at each radius a constant density for the field population as determined in Section~\ref{fieldcont}. By subtracting the field density, a field-decontaminated density profile of the cluster can be produced. An example of the galaxy density as a function of cluster radius is shown in the top panel of Figure~\ref{exampleradprof} for CLJ1226. The average value of the galaxy density of the field is also shown, as well as the field-subtracted radial profile of the cluster.

At any given radius, the total number of galaxies ($T$) is equal to the number of galaxies in the field ($F$) plus the number of galaxies in the cluster ($C$). Therefore, $T = F + C$ and by subtracting the field from the total, we estimated the number of galaxies in the cluster ($C$) at any given radius. From this, we calculate the percentage of galaxies at each radius in the cluster and the field. 
\begin{equation}\label{} \% \; galaxies \; in \; the \; field = F / T\end{equation}
\begin{equation}\label{} \% \; galaxies \; in \; the \; cluster = C / T \end{equation}
\noindent
The resulting values were fit with a polynomial to model the radial probability profile of each cluster. We then determine the probability from 0 (not in the cluster) to 1 (in the cluster) for each galaxy in our image based on its distance from the center of the cluster. An example is shown in the bottom panel of Figure~\ref{exampleradprof}. Based on these radial probability profiles, each galaxy is assigned a weight equal to its cluster membership likelihood. Galaxies with spectroscopic data are assigned a weight of 1 or 0 based on whether their spectroscopic redshift lies within the range of cluster redshift values given in Table~\ref{zsandranges}.


\subsection{Colour Selection and Cluster Membership Probability}
\label{colorprob}

In addition to using radial distance in the cluster to determine probability of cluster membership, we also use galaxy colour to increase cluster membership probability for galaxies with colours matching those of known cluster members at the cluster redshift. The galaxy cluster sample can be roughly divided into three redshift bins at z = 0.5, 0.7, and 0.85. For each of the three redshift groups, we examined the \textit{V-I} colours for the cluster with the most spectroscopically-confirmed cluster members (MACSJ0717 at z $\sim$ 0.5, MACSJ0744 at z $\sim$ 0.7, and MS1054 at z $\sim$ 0.85) and compared the peak of the cluster member galaxy colour distribution with the expected galaxy colours for early type galaxies (E/S0) at these redshifts in \citet{fukugita95}. The observed peak in \textit{V-I} (F555W-F814W) colours of cluster members galaxies in MACSJ0717 is 2.54, which very closely matches the expected galaxy colour of \textit{V-I} = 2.48 given in \citet{fukugita95} for early type galaxies. 

In the case of MS1054, the \textit{V-I} colour we have available is F606W-F775W, which cannot be directly compared with \citet{fukugita95}, where colours are provided in different filters. We therefore use the published values of \citet{tran07}, who find a mean colour of F606W-F775W = 1.61 for red, bright elliptical galaxies in MS1054. This is consistent with the value of 1.69 that we measure for the cluster members. \citet{fukugita95} does not provide galaxy colours at a redshift of 0.7 so a linear extrapolation was fit to the expected \textit{V-I} galaxy colours for elliptical galaxies between a redshift of 0.5 and 0.8. The value of 2.79 we obtain from this fit is consistent with the observed peak value of 2.66 for cluster members in MACSJ0744.

\begin{figure} 
\includegraphics[width=84mm]{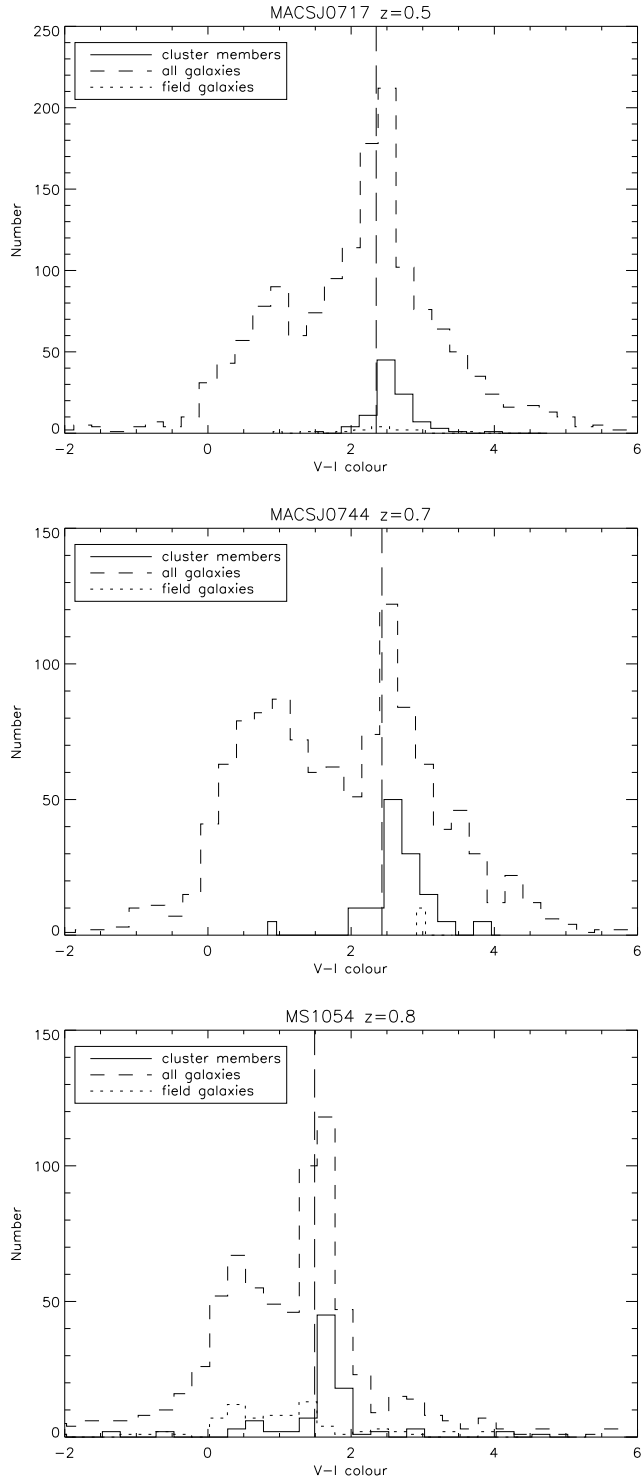}
\caption[\textit{V-I} Colour Distribution of Confirmed Cluster Galaxies]{\textit{V-I} colour distributions of galaxies in MACSJ0717 (top left), MACSJ0744 (top right), and MS1054 (bottom left). The solid histogram shows the spectroscopically-confirmed cluster members and the dotted histogram shows spectroscopically-confirmed non-cluster members. The solid and dotted histograms have been scaled by a factor of 5--10 for visualisation purposes. The vertical dashed line indicates the \textit{V-I} colour limit for galaxies in the cluster. Galaxies redder than this limit have a greater probability of residing in the cluster.}
\label{colorselection}
\end{figure}

\begin{figure} 
\includegraphics[width=84mm]{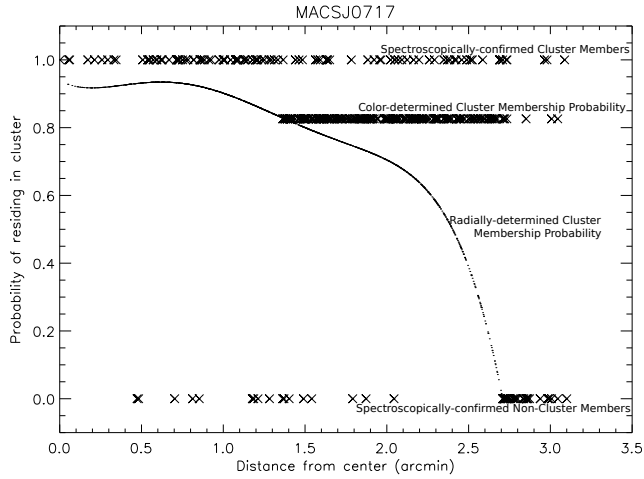}
\caption[Radial Distribution of Galaxies in MACSJ0717]{Radial profile of the cluster MACSJ0717, showing cluster membership probability vs. distance from the center of the cluster with cluster membership probability categories indicated.}
\label{colorprobexample}
\end{figure}

Figure~\ref{colorselection} shows the \textit{V-I} galaxy colour distributions for spectroscopically-confirmed cluster members (solid line), non-cluster members (dotted line), and all galaxies in the cluster image (dashed line) for MACSJ0717, MACSJ0744, and MS1054. Based on the colour distribution of cluster members, we estimate that all galaxies with colours redder than the blue end of this distribution have a high probability of residing in the cluster. We determine the \textit{V-I} colour threshold to be F555W-F814W = 2.36 for clusters at redshift z $\sim$ 0.5, F555W-F814W = 2.43 for clusters at redshift z $\sim$ 0.7, and F606W-F775W = 1.49 for clusters at redshift z $\sim$ 0.85 based on the colour distribution for cluster members in these three representative clusters. In order to determine the probability that should be assigned to galaxies redder than this limit, we examined the radial profiles of cluster members in all clusters and determined the average value of their radially-determined cluster membership probability. This was generally found to be $\sim$80\% for all clusters. We therefore increase the cluster membership probability to 80\% for galaxies redder than the colour thresholds determined here. If a red galaxy had already been assigned a cluster membership probability higher than 80\%, it retained the higher probability value. Figure~\ref{colorprobexample} shows an example of galaxy cluster membership probability values for the cluster MACSJ0717. 

Because the population of confirmed cluster members in these clusters consist of mainly red, early type galaxies, we could not determine a colour-based probability for blue galaxies. Thus, while bluer galaxies did not receive decreased membership probability values, their probabilities are based solely on their radial distance from the center of the cluster or spectroscopic information when available, as statistical information about the blue galaxy population is not easily determined.

 \section{Percentage of AGN in Clusters}
 \label{AGNdiscussion}
 
To determine the percentage of AGN in clusters, we divide the total number of AGN in each cluster by the total number of galaxies in the cluster. First, we compute the percentage for spectroscopically-confirmed cluster members only (where SC stands for "spectroscopically confirmed"):

\begin{equation}\label{agnpercentcm} \% \; AGN_{spec} = \frac{\#\; SC \; cluster \; AGN}{\#\; SC \; cluster \; galaxies}\end{equation}

\noindent
Since spectroscopic coverage varies from cluster to cluster and is sometimes sparse, we also calculate an AGN percentage using all sources in the ACS FOV for each cluster, which covers roughly half of the virial radius. We call this the non-weighted AGN percentage. Finally, we compute our weighted AGN percentage using the cluster membership probabilities described in Sections~\ref{radialprob} and \ref{colorprob} based on radial distance and colour information (where CMP stands for "cluster membership probability): 

\begin{equation}\label{agnpercentweight} \% \; AGN_{weighted} = \frac{\sum(AGN\:CMP)}{\sum(galaxy\:CMP)}\end{equation}

Table~\ref{AGNpercenttable} lists the weighted and non-weighted total percentages of AGN in our galaxy clusters as well as the percentage of AGN detected among spectroscopically-confirmed cluster members. We find that our clusters have a range of 0.8--3.75\% AGN (weighted), with a median value of 2.27 $\pm$ 1.5\%. If we consider confirmed cluster members only, we find that for the 11 clusters with redshift information, 25/530 or 4.7\% of confirmed cluster members show evidence of nuclear activity. Table~\ref{AGNpercenttablebytype} lists the percentage of cluster AGN detected using each of the 3 techniques. In both Table~\ref{AGNpercenttable} and Table~\ref{AGNpercenttablebytype}, a dash (-) indicates that data is not available while an x indicates there are no AGN detected, though data exist. 

We observe that the fraction of AGN among spectroscopically-confirmed cluster galaxies is systematically higher than the fraction observed among field-corrected galaxies. This discrepancy can be resolved by noting that the probability of hosting an AGN increases with host galaxy brightness \citep[e.g.,][]{ho97,sarajedini03}, as well as the fact that any contribution from the AGN's luminosity may serve to increase its host galaxy's brightness. Given that the brightest cluster galaxies are systematically chosen for redshift observations because of their luminosity, we expect to see an increase in the AGN percentage among spectroscopically-confirmed cluster members versus our field-corrected cluster members. We will show in Section~\ref{comp} that when we impose a magnitude limit on our sample, we also observe a higher percentage of AGN. \citet{sarajedini11} observe a similar result, as illustrated in their Figure 9 showing that the fraction of galaxies hosting AGN increases with host galaxy luminosity.

\subsection{Comparison with Field AGN Surveys}
\label{comp}

In \citet{sarajedini11}, AGN candidates were identified in the GOODS fields using the same identification techniques as those used in this study. They identified 85 optically varying galaxies, 259 X-ray sources, and 22 IR power-law sources in these fields. To compare with our cluster survey, we impose several flux limits to the different survey samples. We consider only the sixty-eight X-ray sources in GOODS detectable down to a limit of $\sim$2 x 10$^{-15}$ erg/cm$^{2}$/s, similar to the average depth reached in the X-ray observations of our cluster sample. We further limit the field sample to the redshift range of our clusters (z = 0.4--0.9). With these restrictions, we find 31/1235 galaxies in the GOODS fields host AGN (2.5 $\pm 1.6$\%). This is comparable to, though slightly higher than, the median weighted value we find of 2.27 $\pm$ 1.5\% in clusters. If we further limit our cluster sample to only the 7 in which all 3 AGN selection criteria are possible (i.e., those with mid-IR observations), the AGN percentage is 2.5 $\pm$ 1.6\%, exactly equal to the field AGN percentage.

It is also necessary to ensure that the galaxy magnitude distributions of the GOODS fields and our cluster galaxies are comparable. Since the GOODS survey individual epoch images are shallower than our cluster images (1000s, compared with 2100--3000s), the variability analysis in the GOODS fields extends only to $\sim$24.5, while our cluster photometry extends to \textit{I} $\sim$ 26. We therefore impose a galaxy magnitude limit of \textit{I} = 24.5 on all objects in our cluster sample to further match the GOODS sample. This results in a median weighted AGN percentage among all our clusters of 4.01\%, which rises to 4.94 $\pm$ 2.2\% if we consider only those clusters with IR observations (and thus all three AGN selection techniques are possible). With only the number of known cluster members in these seven clusters, the AGN percentage is 5.5 $\pm$ 2.3\% (20/363). These findings indicate that the number of AGN among galaxies in clusters ($\sim$5\%) is similar or slightly greater than that in the field ($\sim$2.5\%), though only at $\sim$1$\sigma$ significance.

\subsection{Comparison with Cluster X-ray Surveys for AGN}

\citet{martini02} found a lower limit of $\sim$5\% AGN in the cluster A2104 (z = 0.154) with optical counterparts down to an absolute magnitude of R $<$ 20. This corresponds to six X-ray-detected AGN, only one of which shows optical emission lines in its spectra indicative of AGN activity. This is consistent with emission line surveys such as \citet{dressler99}, and suggests that the fraction of AGN in clusters may be higher than previously determined if additional AGN detection techniques are employed.

\citet{martini09} conducted a survey of the luminous AGN population in clusters out to z = 1.3. Their sample includes two of our clusters, MS0451 and MS1054. They required that their sources 1) must have a hard X-ray luminosity L$_{X,H}$ $\ge$ 10$^{43}$ erg/s, 2) the redshift must be within 3 times the cluster's velocity dispersion of the cluster redshift, 3) the source must lie within the projected virial radius of the cluster, and 4) the absolute magnitude of the host galaxy must be M$_{R}$ = M$^{\ast}$(z) + 1. They detect no X-ray AGN in MS0451 meeting this criteria, and when these criteria are applied to our X-ray sources in that cluster, we come to the same result. \citet{martini09} detect one X-ray point source meeting their requirements in the cluster MS1054, but it falls outside of our ACS FOV and therefore was not included in our survey. We also do not detect any other X-ray sources within the ACS FOV that meet these criteria. 

We also compare our results with the X-ray source catalog of \citet{johnson03}, in which the authors identify 47 X-ray point sources in MS1054 down to fluxes similar to those achieved in our data. They find a $\sim$2$\sigma$ excess of X-ray point sources in this particular cluster down to an X-ray flux of 5 x 10$^{-15}$ erg/s/cm$^{2}$, which is consistent with an excess of $\sim$6 AGNs relative to the field. Their survey area is 8.3 x 8.3 arcmin$^{2}$, a factor of 2.5 times larger than the area used in our variability survey per cluster field. Seven of the sources found in \citet{johnson03} fall within our survey FOV and we detect six of these sources, thus confirming their result.

\subsection{Cluster Properties}

\begin{figure}
\centering
\includegraphics[width=84mm]{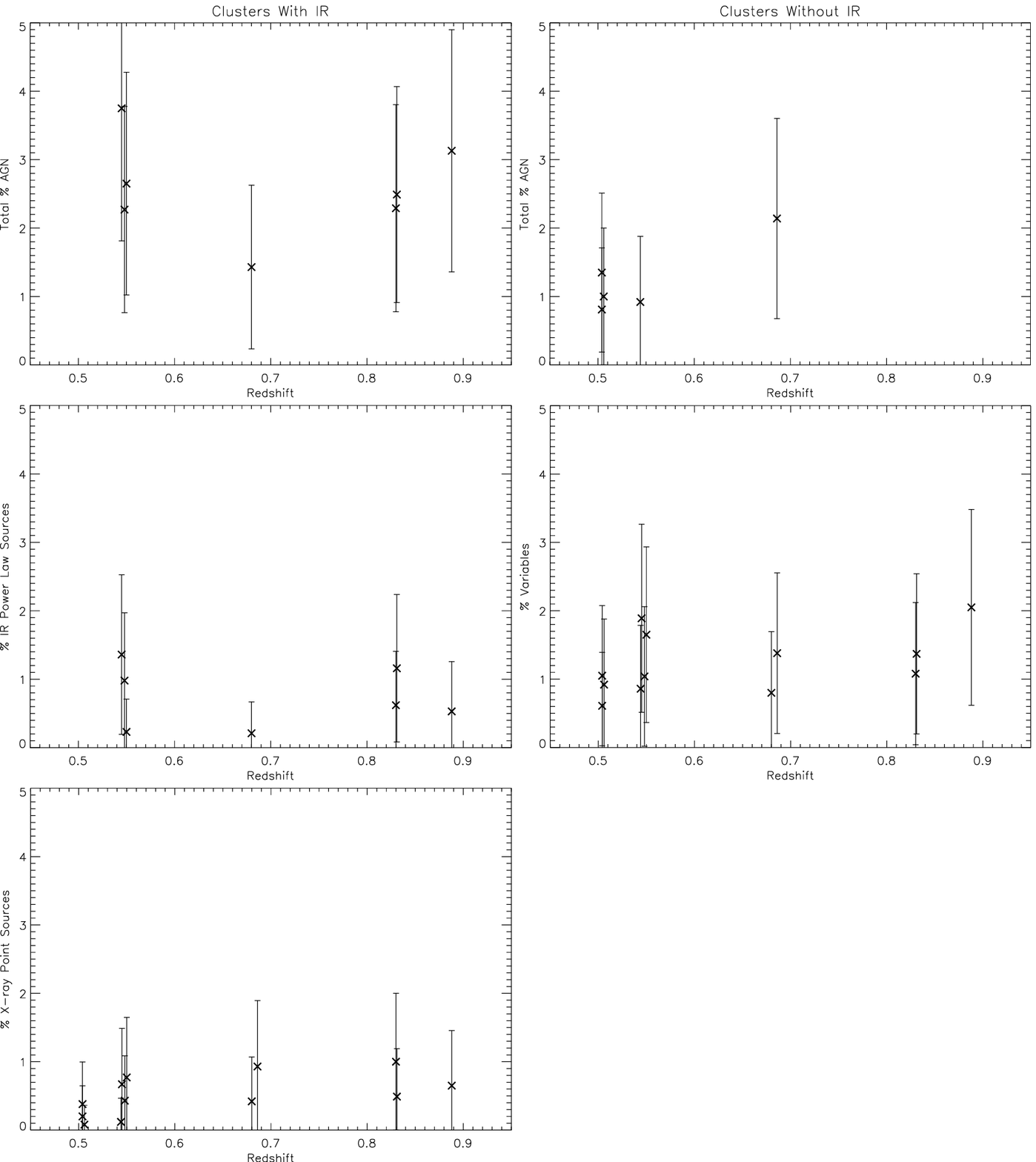}
\caption[AGN \% vs. Cluster Redshift]{AGN \% as a function of cluster redshift. The top two panels show the total percentage of AGN detected; the left panel shows clusters with IR observations and the right panel shows those without. The last two rows show the percentage of IR power law sources, optical variables, and X-ray point sources in each cluster.}
\label{redshift}
\end{figure}

Finally, we investigate whether cluster properties such as redshift, mass, luminosity, velocity dispersion, virial radius, and morphology have any impact on the percentage of AGN. \citet{galametz09} and \citet{martini09} both find evidence for an increase in the number of AGN in clusters with redshift. Figure~\ref{redshift} shows the total weighted percentage of AGN in our clusters as a function of cluster redshift. Over the redshift range of our survey, z = 0.5--0.9, the percentage of AGN appears to be constant. The percentages of variables, X-ray point sources, and IR power-law sources also remain roughly constant regardless of cluster redshift. These results are not inconsistent with the findings of previous studies, as the redshift range covered by our clusters is slightly narrower than that of these studies. This figure suggests that we reach similar levels of completeness in detecting AGN over the redshift range of our cluster sample.

\begin{figure}
\centering
\includegraphics[width=84mm]{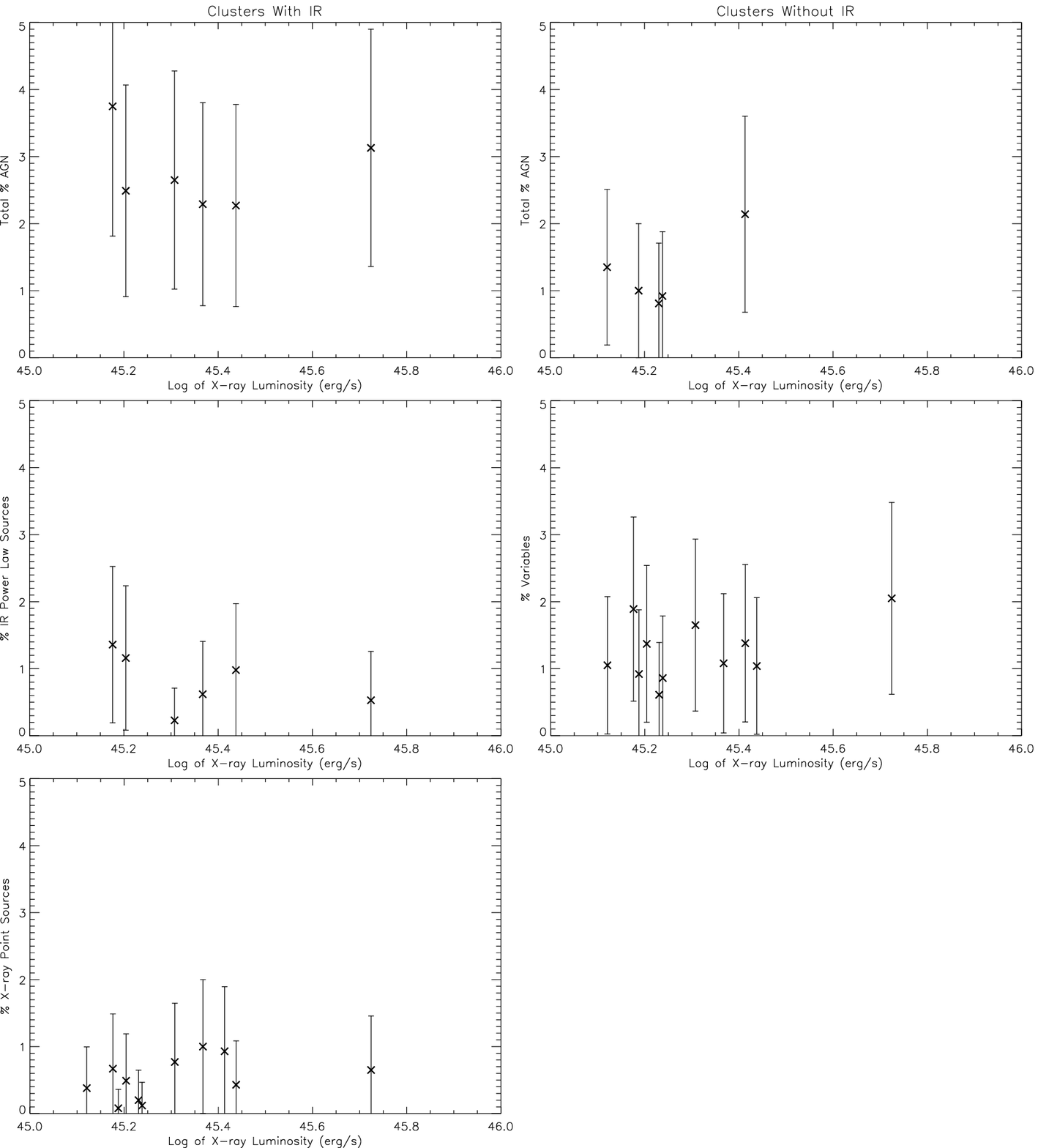}
\caption[AGN \% vs. Cluster X-ray Luminosity]{AGN \% as a function of cluster X-ray Luminosity. The top two panels show the total percentage of AGN detected; the left panel shows clusters with IR observations and the right panel shows those without. The last two rows show the percentage of IR power law sources, optical variables, and X-ray point sources in each cluster.}
\label{luminosity}
\end{figure}

The cluster X-ray luminosity comes from thermal bremsstrahlung emission from hot (T $\sim$ 10$^{8}$ K) intracluster gas bound to the gravitational potential and is therefore related to the cluster mass. As shown in Figure~\ref{luminosity}, there appears to be no change in the total weighted percentage of AGN with the cluster's X-ray luminosity. There also appears to be no significant correlation between the percentage of optical variables, IR power law sources, or X-ray point sources with the X-ray luminosity of the cluster. Thus, the fraction of AGN among galaxies in clusters does not appear to depend on hot gas content over the range sampled by our survey clusters.

\begin{figure}
\centering
\includegraphics[width=84mm]{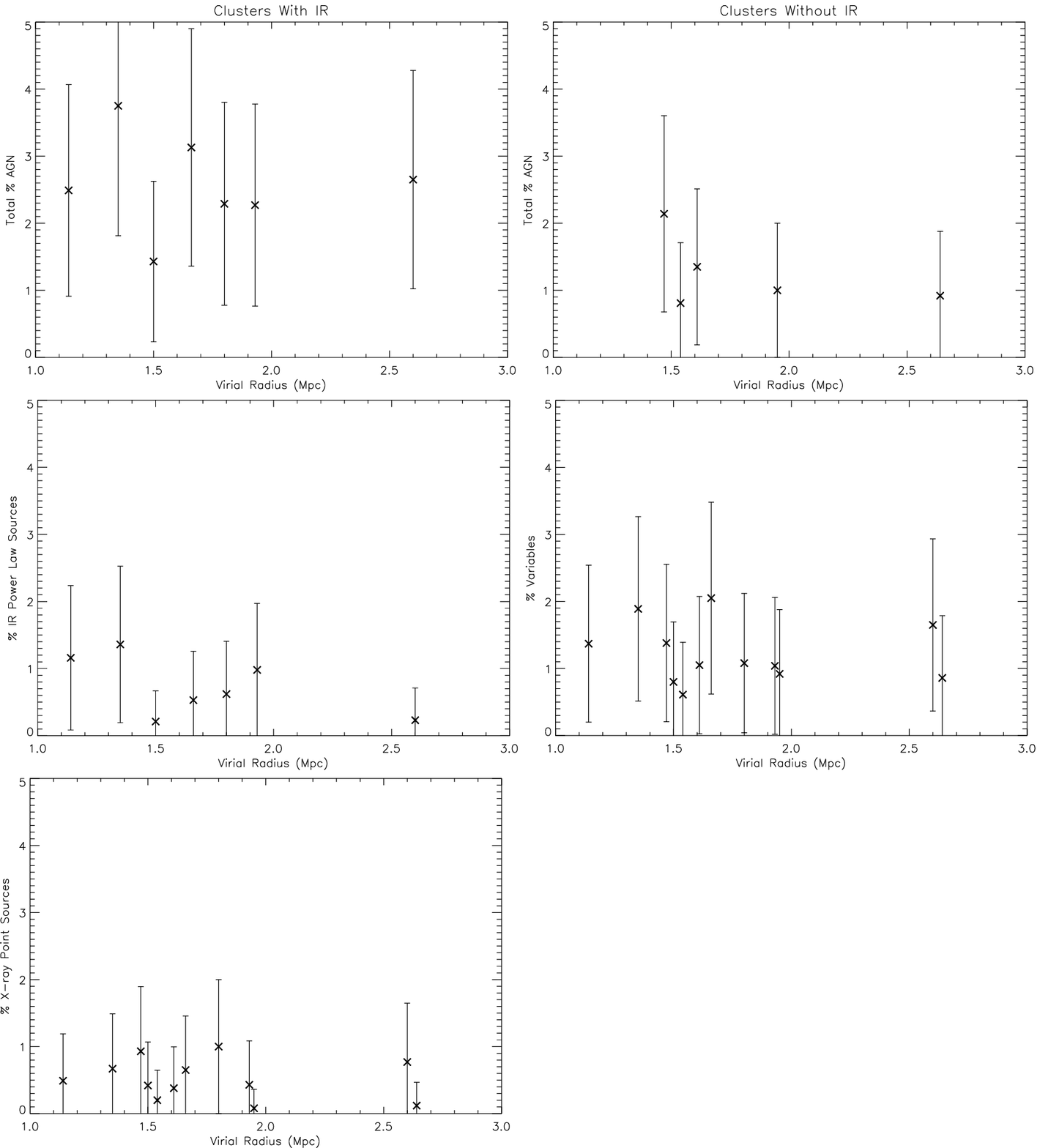}
\caption[AGN \% vs. Cluster Virial Radius]{Percentage of AGN detected as a function of cluster virial radius (Mpc). The top two panels show the total percentage of AGN detected; the left panel shows clusters with IR observations and the right panel shows those without. The last two rows show the percentage of IR power law sources, optical variables, and X-ray point sources in each cluster.}
\label{vr}
\end{figure}

In Figure~\ref{vr} we investigate correlations between the percentage of AGN and the cluster size measured by its virial radius. Though our clusters cover a large range of virial radii, from $\sim$1.1--2.6 Mpc, the total percentage of AGN detected in the clusters remains roughly constant over this range. This is also the case for the optical variables and the X-ray point sources, though the IR power law sources reveal a very slight decline in the percentage of IR sources detected as a function of cluster virial radius. Additionally, we observe no trend between AGN percentage and cluster virial mass, which is not surprising given the correlation between virial mass and virial radius. 

\begin{figure}
\centering
\includegraphics[width=84mm]{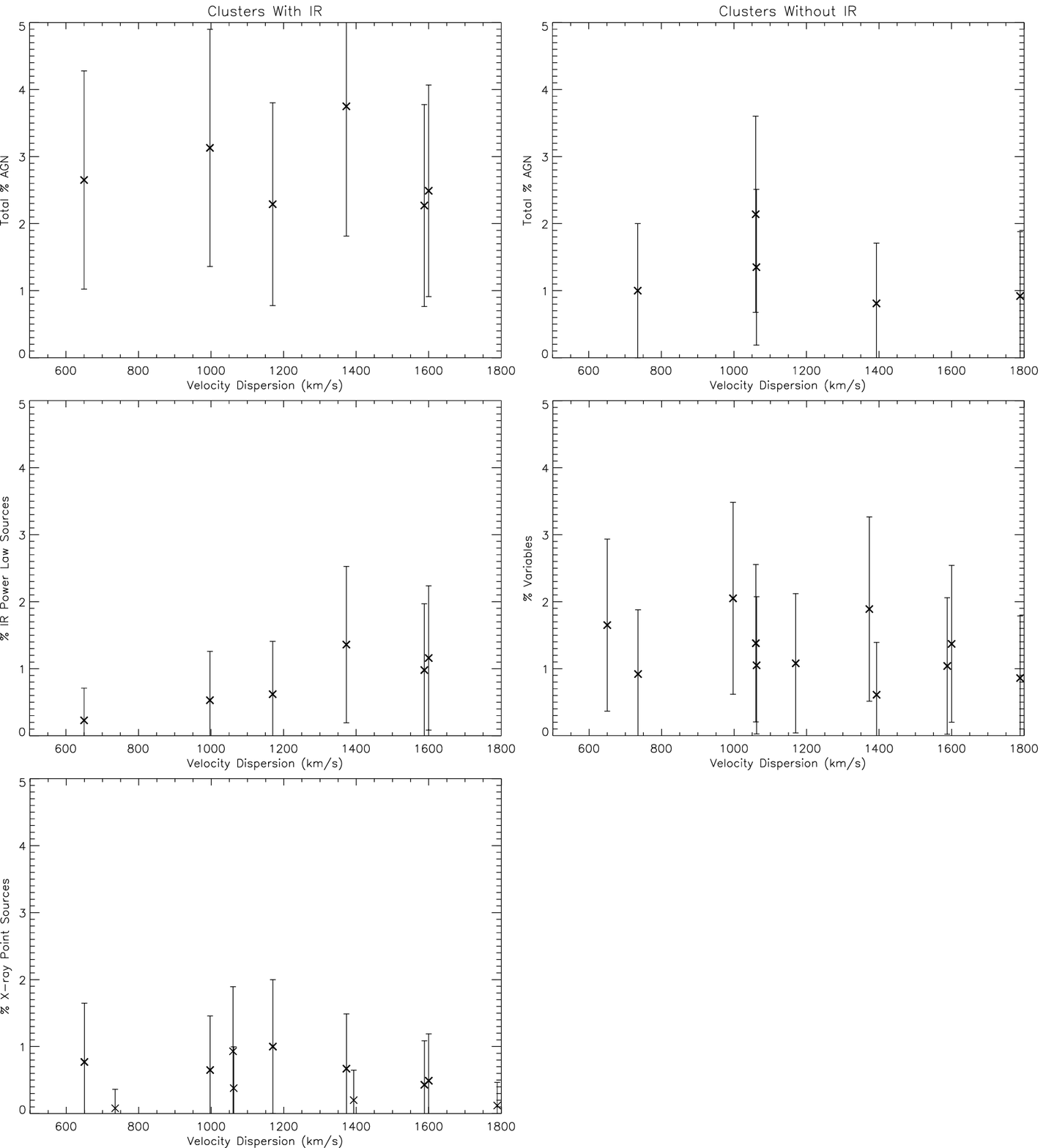}
\caption[AGN \% vs. Cluster Velocity Dispersion]{Percentage of AGN detected as a function of cluster velocity dispersion (km/s). The top two panels show the total percentage of AGN detected; the left panel shows clusters with IR observations and the right panel shows those without. The last two rows show the percentage of IR power law sources, optical variables, and X-ray point sources in each cluster.}
\label{veldisp}
\end{figure}

In Figure~\ref{veldisp}, we look at the percentage of AGN versus the cluster velocity dispersion. There appears to be no significant correlation between the total percentage of AGN and the cluster velocity dispersion, even over our range of 600--1800 km/s. This is also true for optical variables and X-ray point sources, though the IR power law sources show a slight increase in the AGN fraction with increasing velocity dispersion.

\citet{ebeling07} assigns each of the MACS clusters in our sample a morphology code, dependent on the agreement between the X-ray and optical emission in the cluster, as well as signs of disturbances in the cluster substructure. Their morphology codes are assigned based on the following criteria: \\

\noindent
1: Relaxed (pronounced cool core, perfect alignment of X-ray peak and single cD galaxy) \\
2: Semi-Relaxed (good optical/X-ray alignment, concentric contours) \\
3: Semi-Disturbed (nonconcentric contours, obvious small-scale substructure) \\
4: Disturbed (poor optical/X-ray alignment, multiple peaks, no cD galaxy) \\

\noindent
Based on these definitions, they gave the galaxy clusters that appear in our sample the following morphology classifications: \\

\noindent
MACSJ0257: 2 (Semi-Relaxed) \\
MACSJ0717: 4 (Disturbed) \\
MACSJ0744: 2 (Semi-Relaxed) \\
MACSJ0911: 4 (Disturbed) \\
MACSJ1149: 4 (Disturbed) \\
MACSJ1423: 1 (Relaxed) \\
MACSJ2214: 2 (Semi-Relaxed) \\

\noindent 
Using the classification criteria of \citet{ebeling07}, we assign a morphology code to each of our remaining clusters: \\

\noindent
CL0152: 4 (Disturbed) \\
CLJ1226: 1 (Relaxed) \\
MS0451: 1 (Relaxed) \\
MS1054: 3 (Semi-Disturbed) \\
SDSS1004: 1 (Relaxed) \\

\noindent
With only a small number of clusters in each of these morphological groups, we combine the relaxed and semi-relaxed clusters (7 clusters), and also the disturbed and semi-disturbed clusters (5 clusters). A comparison of the median value of the weighted percentage of AGN shows that both groups have very similar percentages of AGN: the relaxed clusters have a median value of 2.13\% and the disturbed clusters have a median value of 2.27\%. If we exclude clusters with no IR observations, we have 3 clusters in the disturbed group and 4 clusters in the relaxed group. We find that the disturbed clusters have a median percentage of 2.29\% cluster AGN, while the relaxed clusters have a median AGN percentage of 3.13\%. Thus, we do not find evidence that the cluster dynamical state has an impact on the number of AGN detected in our sample of clusters.

\subsection{Summary: AGN Percentage}

Comparing our results with the multi-wavelength GOODS AGN field survey of \citet{sarajedini11}, we find that the percentage of AGN is similar to or slightly enhanced in clusters relative to the field when considering AGN identified via optical variability, X-ray emission, or mid-IR power-law SEDs. We also find that the number of X-ray-detected AGN in our survey is consistent with other surveys for X-ray point sources in galaxy clusters, which have reported an enhancement of X-ray sources among some galaxy clusters relative to the field. These two findings may at first appear to be inconsistent. Are the number of AGN in clusters more than or the same as that found in the field? The answer seems to depend strongly on completeness issues and correcting for galaxy magnitude limits as well as X-ray flux limits in the surveys being compared. In any case, it seems clear that AGN activity is in no way inhibited in the cluster environment. 

This is evidence that galaxies are still able to fuel accretion onto their supermassive black holes, even in denser environments. \citet{martini04} point out that while major mergers may be perhaps the only reasonable candidate to trigger and sustain luminous AGN, lower luminosity AGN such as those identified by our survey may have significantly more physical processes capable of fuelling a central supermassive black hole, including bar structures, minor mergers, galaxy harassment, and stellar mass loss -- all of which still play a significant role in a galaxy cluster environment. If this is in fact the case, there may be a comparable number of lower luminosity AGN in clusters and in the field, which is consistent with our results, as optical variability and IR power law detection are likely to pick out lower luminosity AGN. This is also consistent with the picture that more luminous AGN may be less common in denser environments \citep[e.g.,][]{kauffmann04, pb06}, as we find no luminous X-ray sources in MS0451 or MS1054, consistent with the survey of \citet{martini09}. 

We find no obvious trends between the AGN fraction in clusters and various cluster properties, including mass, X-ray luminosity, virial radius, velocity dispersion, redshift and morphology. We do note that many of these properties are interrelated, and in general the clusters in our sample cover a small range of mass and X-ray luminosity. We find that the number of AGN we detect is roughly constant regardless of redshift, confirming that we are able to reach similar levels of completeness over the range of cluster redshifts in our sample. Finally, we do not observe a significant link between cluster morphology and the number of AGN we detect, finding similar percentages of cluster AGN in both disturbed and relaxed clusters.

\section{Conclusions}
\label{conclusions}

We have explored several issues concerning the AGN population in dense cluster environments. We analysed 12 galaxy clusters at redshifts 0.5 $<$ z $<$ 0.9 to determine the AGN fraction and address the issue of AGN fuelling in massive galaxy clusters. We compile, for the first time, a catalog of cluster AGN candidates using a combination of three detection techniques: optical variability, X-ray point source detection, and mid-IR power-law SEDs. 

We identify 178 optical variables among the galaxies in our cluster sample using 2--3 epochs of ACS imaging, an average of 15 variables per cluster. Ninety (51\%) of these variables have $>$4$\sigma$ significance. We find that in total, 1.1\% of all galaxies surveyed display nuclear optical variability in galaxies to \textit{I}$_{nuc}$ = 27, corresponding to an absolute magnitude of M$_{I}$ $\sim$ -15.3 at a redshift of 0.5 and M$_{I}$ $\sim$ -16.8 at a redshift of 0.9.

We find 74 X-ray point sources down to a full band flux of $\sim$7 x 10$^{-16}$ erg/cm$^{2}$/s using Chandra X-ray imaging, an average of 6 X-ray point sources per cluster. This flux corresponds to an X-ray luminosity of $\sim$6 x 10$^{40}$ erg/s at a redshift of 0.5 and $\sim$3 x 10$^{42}$ erg/s at a redshift of 0.9. Most of the point sources have hardness ratios on the soft end of the distribution, with 46 sources (67\%) having F$_{X}$(2--8 keV)$/$F$_{X}$(0.5--2 keV) $\le$ 5. Seventy percent of our X-ray sources lie within $log$(F$_{X}$/F$_{opt}$) = $\pm$1, consistent with the presence of an AGN. 

Spitzer IRAC data is available for 7 clusters, in which we identify a total of 64 IR power law sources. This is an average of 9 IR power law sources per cluster. Of these objects, 44\% show BLAGN-like mid-IR SEDs, while 31\% have steeper, NLAGN-like SEDs. We find that 87--100\% of IR power law sources are in the Lacy AGN wedge, while 13--16\% of optical variables and 38--47\% of X-ray point sources lie in the Lacy wedge. This indicates that the majority of optically variable AGN and about half of X-ray-selected AGN are not dominated by the AGN light in the mid-IR. 

In total, we find an average of 25 AGN candidates per cluster with a range of 12--49 per cluster over the sample. We identify 50 X-ray point sources and 48 mid-IR power law sources with optical counterparts, and find that 4\% of mid-IR power law sources and 24\% of X-ray point sources are detected as optical variables. Only seven percent of variables are also detected either through X-ray emission or as mid-IR power law sources. Among the X-ray and IR objects we find that 9\% of IR power law sources are also detected via X-rays, while 8\% of X-ray point sources also show a power law SED in the mid-IR. 

A primary goal of our work is to calculate the percentage of cluster AGN, taking into account the cluster membership probability for each galaxy. We find that our clusters have a range of $\sim$1--4\% AGN with a median value of 2.3 $\pm$ 1.5\%. We compare the percentage of cluster AGN with the percentage of AGN among field galaxies in the GOODS field detected via optical variability, X-rays, or mid-IR SED fitting. Within the same redshift and X-ray flux limits as our cluster data, 2.5 $\pm 1.6$\% of field galaxies are found to host AGN. Applying a galaxy magnitude limit to our cluster data to match the shallower exposure times in GOODS, we find a median weighted AGN percentage among the clusters of 4.94 $\pm$ 2.2\%. Thus, we find that the number of AGN among galaxies in clusters is essentially the same as that in the field. We find no obvious trends among cluster properties and the percentage of AGN detected, though in most cases our clusters cover a relatively small range of these properties. We also do not find a significant difference in the percentage of cluster galaxies that host AGN as a function of cluster dynamical state or morphology.

Our results confirm earlier indications that galaxies are able to fuel accretion onto the central supermassive black hole even in denser cluster environments. While major mergers may be a strong candidate to trigger and sustain luminous AGN, lower luminosity AGN, such as many of those identified by our survey, may be triggered or sustained through a greater variety of physical processes including bar structures, minor mergers, galaxy harassment, and stellar mass loss -- all of which still play a significant role in the cluster environment. 

In a future paper we will present the radial distribution of AGN among cluster galaxies to examine how local environment affects the spatial distribution of AGN within the cluster. We will also explore the properties of the AGN host galaxies in our survey to investigate the link between AGN activity and the evolution of galaxies. Using our sample, we will examine the relationship between the presence of an AGN, the host galaxy, and the cluster environment.


\bibliography{references}


\clearpage
\newpage

\clearpage

\clearpage

\clearpage

\clearpage

\clearpage

\clearpage

\clearpage

\clearpage

\clearpage
\newpage

\begin{table*}
 \centering
 \begin{minipage}{140mm}
 \caption[Cluster Observations]{Cluster Observations \label{clustersgen}}
\begin{tabular}{lccrcc}
 \hline
 \multicolumn{1}{c}{Cluster} &
 \multicolumn{1}{c}{Redshift} &
 \multicolumn{1}{c}{RA} &
 \multicolumn{1}{c}{Dec} &
 \multicolumn{1}{c}{\#ACS Epochs} &
 \multicolumn{1}{c}{Obs. w$/$Spitzer} \\
\hline
$CL0152-1357$       &  0.831  &  01:52:43.00  & -13:57:20.00 & 2  & Y  \\
$CLJ1226.9+3332$   &  0.888  &  12:26:58.21  & 33:32:49.4     & 2  & Y  \\
$MACSJ0257-2325$ &  0.506  &   02:57:08.83  & -23:26:03.3    & 3  & -  \\
$MACSJ0717+3745$ &  0.548  &  07:17:32.93 & 37:45:05.4     & 3  & Y  \\
$MACSJ0744+3297$ &  0.686  &  07:44:52.58 & 39:27:26.7     & 3  & -  \\
$MACSJ0911+1746$ &  0.505  &  09:11:11.18 & 17:46:34.8     & 3  & -  \\
$MACSJ1149+2223$ &  0.544  &  11:49:35.51 & 22:24:04.2     & 2  & -  \\
$MACSJ1423+2404$ &  0.545  &  14:23:48.60 & 24:04:49.1     & 2  & Y  \\
$MACSJ2214-1359$  &  0.504  &  22:14:57.34 & -14:00:12.2   & 2  & -  \\
$MS0451.6-0305$     &  0.550  &  04:54:10.48 & -03:01:38.5   & 2  & Y  \\
$MS1054.4-0321$     &  0.830  & 10:57:00.20 & -03:37:27.0   & 2  & Y  \\
$SDSS1004+41$       &  0.680  &  10:04:34.72 & 41:12:44.98   & 3  & Y  \\
\hline
\end{tabular}
\end{minipage}
\end{table*}

\begin{table*}
 \centering
 \begin{minipage}{140mm}
\caption[Galaxy Cluster Properties]{Galaxy Cluster Properties \label{clusterprops}}
\begin{tabular}{lccccc}
\hline
\multicolumn{1}{c}{Cluster} &
\multicolumn{1}{c}{L$_{X} (10^{44} erg/s$)} &
\multicolumn{1}{c}{M$_{200} (10^{15} M_{\odot})$} &
\multicolumn{1}{c}{r$_{v}$ (Mpc)} &
\multicolumn{1}{c}{$\sigma$ (km/s)} \\
\hline
$CL0152-1357$            & 16    & 0.45 & 1.14                    & 1600 \\
$CLJ1226.9+3332$        & 53    & 1.4   & 1.66                    & 997 \\
$MACS J0257.6-2209$  & 15.4 & 1.41\footnotemark[1] & 	1.95 $\pm$ 0.42 & 735 \\
$MACS J0717.5+3745$ & 27.4 & 2.75\footnotemark[1] & 	1.93 $\pm$ 0.28 & 1588 \\
$MACS J0744.8+3927$ & 25.9 & 0.87\footnotemark[1] & 	1.47 $\pm$ 0.18 & 1060 \\
$MACS J0911.2+1746$ & 13.2 & 0.84\footnotemark[1] & 	1.61 $\pm$ 0.50 & 1062 \\
$MACS J1149.5+2223$ & 17.3 & 1.13\footnotemark[1] & 	2.64 $\pm$ 0.14 & 1789 \\
$MACS J1423.8+2404$ & 15.0 & 0.78\footnotemark[1] & 	1.35 $\pm$ 0.19 & 1373 \\
$MACS J2214.9-1359$ & 17.0 & 1.22\footnotemark[1] & 	1.54 $\pm$ 0.16 & 1393 \\
$MS0451.6-0305$         & 20.3  & 1.4  &  2.6                      & 650 \\
$MS1054.4-0321$         & 23.3  & 1.1  &  1.8                      & 1170 \\
$SDSS1004+41$           & -        & 0.42 & 1.35                      & - \\	
\hline
\end{tabular}
\footnotetext{$^{1}$ The value listed for M$_{200}$ corresponds to M$_{X,200}$ derived from X-ray gas as measured by \citet{barrett06}}
\end{minipage}
\end{table*}

\begin{table*}
 \centering
 \begin{minipage}{140mm}
\caption[ACS Observations]{ACS Observations \label{acsobs}}
\begin{tabular}{lccc}
\hline
\multicolumn{1}{c}{Cluster} &
\multicolumn{1}{c}{\#ACS Epochs} &
\multicolumn{1}{c}{Filter(s)} &
\multicolumn{1}{c}{Total I-band exp. time (s)} \\
\hline
$CL0152-1357$   &   2  & F775W,F625W       & 4252 \\
$CLJ1226.9+3332$ &   2  & F814W       & 4296 \\
$MACSJ0257-2325$ &   3  & F814W,F555W & 8858 \\
$MACSJ0717+3745$ &   3  & F814W,F555W & 8893 \\
$MACSJ0744+3297$ &   3  & F814W,F555W & 8893 \\
$MACSJ0911+1746$ &   3  & F814W,F555W & 8825 \\
$MACSJ1149+2223$ &   2  & F814W,F555W & 6774 \\
$MACSJ1423+2404$ &   2  & F814W,F555W & 6774 \\
$MACSJ2214-1359$ &   2  & F814W,F555W & 6640 \\
$MS0451.6-0305$  &   2  & F814W,F555W       & 4198 \\
$MS1054.4-0321$  &   2  & F775W,F606W       & 4232 \\
$SDSS1004+41$    &   3  & F814W,F555W & 9492 \\
\hline
\end{tabular}
\end{minipage}
\end{table*}

\clearpage

\begin{center}
\begin{longtable}{rrccr}
\caption[Optical Variables: 2 Epochs]{Optical Variables: 2 Epochs \label{2evar}} \\
\hline
\multicolumn{1}{c}{RA} &
\multicolumn{1}{c}{Dec} &
\multicolumn{1}{c}{Magnitude} &
\multicolumn{1}{c}{Magnitude Difference} &
\multicolumn{1}{c}{$\sigma\sigma^{\ast}$} \\
\multicolumn{1}{c}{} &
\multicolumn{1}{c}{} &
\multicolumn{1}{c}{in Epoch 1} &
\multicolumn{1}{c}{$\Delta$$mag_{E1}-mag_{E2}$} &
\multicolumn{1}{c}{} \\
\hline
\endfirsthead
{\normalsize{\tablename\ \thetable{}. Continued}} \\
\hline
\multicolumn{1}{c}{RA} &
\multicolumn{1}{c}{Dec} &
\multicolumn{1}{c}{Magnitude} &
\multicolumn{1}{c}{Magnitude Difference} &
\multicolumn{1}{c}{$\sigma\sigma^{\ast}$} \\
\multicolumn{1}{c}{} &
\multicolumn{1}{c}{} &
\multicolumn{1}{c}{in Epoch 1} &
\multicolumn{1}{c}{$\Delta$$mag_{E1}-mag_{E2}$} &
\multicolumn{1}{c}{} \\
\hline
\endhead
\hline
\multicolumn{3}{l}{{Continued on Next Page\ldots}} \\
\endfoot
\hline
\endlastfoot
{\bf Group 2} \\
\hline
\multicolumn{1}{l}{\bf MACSJ1149} \\
177.3912546 &  22.3741796 & 	   19.761  & 	0.062	 &  4.11      \\
177.3727730 &  22.4131815 & 	   22.169  & 	0.124	 &  8.22      \\
177.3853783 &  22.4158100 & 	   22.178  & 	0.062	 &  4.11      \\
177.4129531 &  22.3912347 & 	   22.684  & 	0.080	 &  5.30      \\
177.4037943 &  22.4045812 & 	   23.072  & 	0.046	 &  3.05	  \\
177.3759224 &  22.4014957 & 	   23.360  & 	0.223	 & 14.78	   \\
\end{longtable}
\end{center}

\clearpage
\newpage

\begin{center}
\begin{longtable}{rrccr}
\caption[Optical Variables: 3 Epochs]{Optical Variables: 3 Epochs \label{3evar}} \\
\hline
\multicolumn{1}{c}{RA} &
\multicolumn{1}{c}{Dec} &
\multicolumn{1}{c}{Average} &
\multicolumn{1}{c}{Standard} &
\multicolumn{1}{c}{$\sigma$} \\
\multicolumn{1}{c}{} &
\multicolumn{1}{c}{} &
\multicolumn{1}{c}{Magnitude} &
\multicolumn{1}{c}{Deviation} &
\multicolumn{1}{c}{} \\
\hline
\endfirsthead
{\normalsize{\tablename\ \thetable{}. Continued}} \\
\hline
\multicolumn{1}{c}{RA} &
\multicolumn{1}{c}{Dec} &
\multicolumn{1}{c}{Average} &
\multicolumn{1}{c}{Standard} &
\multicolumn{1}{c}{$\sigma$} \\
\multicolumn{1}{c}{} &
\multicolumn{1}{c}{} &
\multicolumn{1}{c}{Magnitude} &
\multicolumn{1}{c}{Deviation} &
\multicolumn{1}{c}{} \\
\hline
\endhead
\hline
\multicolumn{3}{l}{{Continued on Next Page\ldots}} \\
\endfoot
\hline
\endlastfoot
\multicolumn{1}{l}{\bf MACSJ0257} \\
44.3055608 &  -23.4389531  &   21.270	 &  0.244  &   19.39	  \\
44.2958934 &  -23.4440126  &   22.429	 &  0.082  & 	6.51	  \\
44.3086292 &  -23.4349091  &   22.914	 &  0.077  & 	5.88	  \\
44.2831760 &  -23.4615253  &   23.837	 &  0.066  & 	5.01	  \\
44.2698898 &  -23.4558108  &   23.851	 &  0.045  & 	3.42	  \\
44.2980084 &  -23.4009620  &   24.237	 &  0.047  & 	3.40	  \\
44.2826153 &  -23.4625286  &   24.504	 &  0.045  & 	3.02	  \\
\end{longtable}
\end{center}

\clearpage
\newpage

\begin{table*}
 \centering
 \begin{minipage}{140mm}
\caption[X-ray Observations]{X-ray Observations \label{xobs}}
\begin{tabular}{lcc}
\hline
\multicolumn{1}{c}{Cluster} &
\multicolumn{1}{c}{Obs IDs} &
\multicolumn{1}{c}{Effective Exp. Time (ks)} \\
\hline
$CLJ0152-1357$       &   913                           &	34.79  \\
$CLJ1226.9+3332$   &   3180, 5014             &	  56.6   \\
$MACSJ0257-2325$ &   1654, 3518             &	36.27  \\
$MACSJ0717+3745$ &   1655, 4200            &	74.98  \\
$MACSJ0744+3297$ &   3197, 3585, 6111 &	86.49  \\
$MACSJ0911+1746$ &   3587, 5012           &	39.11  \\
$MACSJ1149+2223$ &   1656, 3589            &	36.04  \\
$MACSJ1423+2404$ &   1657, 4195            &	131.65 \\
$MACSJ2214-1359$  &   3259, 5011            &	35.07  \\
$MS0451.6-0305$      &   529, 902                &	84.59  \\
$MS1054.4-0321$      &   512                        &	55.81  \\
$SDSS1004+41$        &   5794                      &	73.89  \\
\hline
\end{tabular}
\end{minipage}
\end{table*}

\clearpage
\newpage

\begin{landscape}
\begin{ThreePartTable}
\small
\tabcolsep=1.0mm
\begin{longtable}{cccccccccr}
\caption[X-ray Point Sources]{X-ray Point Sources \label{xpsprop}} \\
28.1822599 &    -13.9837439 &      119.6  &   2.50  &  2.13  &  3.03 &  83.13 &  70.71 & 100.45 & 0.704  \kill
\hline
\multicolumn{1}{c}{RA} &
\multicolumn{1}{c}{Dec} &
\multicolumn{1}{c}{Source} &
\multicolumn{1}{c}{Full Band Flux} &
\multicolumn{1}{c}{Hard Band Flux} &
\multicolumn{1}{c}{Soft Band Flux} &
\multicolumn{1}{c}{Full Band} &
\multicolumn{1}{c}{Hard Band} &
\multicolumn{1}{c}{Soft Band} &
\multicolumn{1}{c}{Hardness} \\
\multicolumn{1}{c}{} &
\multicolumn{1}{c}{} &
\multicolumn{1}{c}{Significance} & 
\multicolumn{1}{c}{(10$^{-14}$ erg/cm$^{2}$/s)} & 
\multicolumn{1}{c}{(10$^{-14}$ erg/cm$^{2}$/s)} & 
\multicolumn{1}{c}{(10$^{-14}$ erg/cm$^{2}$/s)} & 
\multicolumn{1}{c}{Luminosity} & 
\multicolumn{1}{c}{Luminosity} & 
\multicolumn{1}{c}{Luminosity} & 
\multicolumn{1}{c}{Ratio} \\ 
\multicolumn{1}{c}{} &
\multicolumn{1}{c}{} &
\multicolumn{1}{c}{} & 
\multicolumn{1}{c}{} & 
\multicolumn{1}{c}{} & 
\multicolumn{1}{c}{} & 
\multicolumn{1}{c}{(10$^{42}$ erg/s)} & 
\multicolumn{1}{c}{(10$^{42}$ erg/s)} & 
\multicolumn{1}{c}{(10$^{42}$ erg/s)} & 
\multicolumn{1}{c}{} \\ 
\hline
\endfirsthead
{\normalsize{\tablename\ \thetable{}. Continued}} \\
\hline
\multicolumn{1}{c}{RA} &
\multicolumn{1}{c}{Dec} &
\multicolumn{1}{c}{Source} &
\multicolumn{1}{c}{Full Band Flux} &
\multicolumn{1}{c}{Hard Band Flux} &
\multicolumn{1}{c}{Soft Band Flux} &
\multicolumn{1}{c}{Full Band} &
\multicolumn{1}{c}{Hard Band} &
\multicolumn{1}{c}{Soft Band} &
\multicolumn{1}{c}{Hardness} \\
\multicolumn{1}{c}{} &
\multicolumn{1}{c}{} &
\multicolumn{1}{c}{Significance} & 
\multicolumn{1}{c}{(10$^{-14}$ erg/cm$^{2}$/s)} & 
\multicolumn{1}{c}{(10$^{-14}$ erg/cm$^{2}$/s)} & 
\multicolumn{1}{c}{(10$^{-14}$ erg/cm$^{2}$/s)} & 
\multicolumn{1}{c}{Luminosity} & 
\multicolumn{1}{c}{Luminosity} & 
\multicolumn{1}{c}{Luminosity} & 
\multicolumn{1}{c}{Ratio} \\ 
\multicolumn{1}{c}{} &
\multicolumn{1}{c}{} &
\multicolumn{1}{c}{} & 
\multicolumn{1}{c}{} & 
\multicolumn{1}{c}{} & 
\multicolumn{1}{c}{} & 
\multicolumn{1}{c}{(10$^{42}$ erg/s)} & 
\multicolumn{1}{c}{(10$^{42}$ erg/s)} & 
\multicolumn{1}{c}{(10$^{42}$ erg/s)} & 
\multicolumn{1}{c}{} \\ 
\hline
\endhead
\hline
\multicolumn{3}{l}{{Continued on Next Page\ldots}} \\
\endfoot
\hline
\endlastfoot
\multicolumn{1}{l}{\bf CL0152} \\
28.1822599 &    -13.9837439 &      119.6  &   2.50  &  2.13  &  3.03 &  83.13 &  70.71 & 100.45 & 0.704     \\
28.1658054 &    -13.9614279 &       99.7  &   2.97  &  4.59  &  1.27 &  98.65 & 152.55 &  42.01 & 3.631     \\
28.2054645 &    -13.9494400 &        2.9  &   0.02  &  0.39  &  0.06 &   0.72 &  13.00 &   1.94 & 6.713     \\
28.1794661 &    -13.9222264 &        3.4  &   0.05  &  0.001 &  0.05 &   1.62 &   0.05 &   1.63 & 0.029     \\
\multicolumn{1}{l}{\bf CLJ1226} \\						        
186.7745191   &   33.5371130 &        39.1  &   1.54  &  0.96  &  0.51 &  60.44 &  37.40    &  19.99 & 1.871  \\
186.7306779   &   33.5619212 &         4.5  &   0.07  &  0.02  &  0.04 &   2.59 &   0.83    &   1.59 & 0.522  \\
\hline
\end{longtable}
\end{ThreePartTable}
\end{landscape}
\clearpage
\newpage

\begin{table*}
 \centering
 \begin{minipage}{140mm}
\caption[Rest-Frame X-ray Fitting Parameters]{Rest-Frame X-ray Fitting Parameters \label{xrayfit}}
\begin{tabular}{lccc}
\hline
\multicolumn{1}{c}{Cluster} &
\multicolumn{1}{c}{Redshift} &
\multicolumn{1}{c}{n$_{H}$(10$^{20} cm^{-2}$)} &
\multicolumn{1}{c}{D$_{L}$ (10$^{3}$ Mpc)} \\
\hline
$CLJ0152-1357$       &  0.831  & 1.58 & 5.2675 \\
$CLJ1226.9+3332$   &  0.888  & 1.37 & 5.7207 \\
$MACSJ0257-2325$ &  0.506  & 2.16 & 2.8636 \\
$MACSJ0717+3745$ &  0.548  & 7.11 & 3.1546 \\
$MACSJ0744+3297$ &  0.686  & 5.62 & 4.1540 \\
$MACSJ0911+1746$ &  0.504  & 3.64 & 2.8499 \\
$MACSJ1149+2223$ &  0.544  & 2.21 & 3.1266 \\
$MACSJ1423+2404$ &  0.545  & 2.28 & 3.1336 \\
$MACSJ2214-1359$  &  0.504  & 3.30 & 2.8499 \\
$MS0451.6-0305$     &  0.550  & 5.03 & 3.1686 \\
$MS1054.4-0321$     &  0.830  & 3.62 & 5.2596 \\
$SDSS1004+41$       &  0.680  & 1.11 & 4.1092 \\
\hline
\end{tabular}
\end{minipage}
\end{table*}

\clearpage
\newpage

\small
\begin{longtable}{ccrrrrcc}
\caption[IR Power Law Galaxies]{IR Power Law Galaxies \label{plawprop}} \\
28.1657731 &	-13.9614409  & 316.814 $\pm$ 0.028 & 341.394 $\pm$ 0.034 & 425.256 $\pm$ 0.205 & 552.300 $\pm$ 0.157 &  -0.534 $\pm$ 0.000  \kill
\hline
\multicolumn{1}{c}{RA} &
\multicolumn{1}{c}{Dec} &
\multicolumn{1}{c}{Channel 1} &
\multicolumn{1}{c}{Channel 2} &
\multicolumn{1}{c}{Channel 3} &
\multicolumn{1}{c}{Channel 4} &
\multicolumn{1}{c}{$\alpha$} \\
\multicolumn{1}{c}{} &
\multicolumn{1}{c}{} &
\multicolumn{1}{c}{Flux ($\mu$Jy)} &
\multicolumn{1}{c}{Flux ($\mu$Jy)} &
\multicolumn{1}{c}{Flux ($\mu$Jy)} &
\multicolumn{1}{c}{Flux ($\mu$Jy)} &
\multicolumn{1}{c}{} \\
\hline
\endfirsthead
{\normalsize{\tablename\ \thetable{}. Continued}} \\
\hline
\multicolumn{1}{c}{RA} &
\multicolumn{1}{c}{Dec} &
\multicolumn{1}{c}{Channel 1} &
\multicolumn{1}{c}{Channel 2} &
\multicolumn{1}{c}{Channel 3} &
\multicolumn{1}{c}{Channel 4} &
\multicolumn{1}{c}{$\alpha$} \\
\multicolumn{1}{c}{} &
\multicolumn{1}{c}{} &
\multicolumn{1}{c}{Flux ($\mu$Jy)} &
\multicolumn{1}{c}{Flux ($\mu$Jy)} &
\multicolumn{1}{c}{Flux ($\mu$Jy)} &
\multicolumn{1}{c}{Flux ($\mu$Jy)} &
\multicolumn{1}{c}{} \\
\hline
\endhead
\hline
\multicolumn{3}{l}{{Continued on Next Page\ldots}} \\
\endfoot
\hline
\endlastfoot
\multicolumn{1}{l}{\bf CL0152} \\
28.1822554 &	-13.9837263  &  84.382 $\pm$ 0.03 &  88.462 $\pm$ 0.03 &  96.722 $\pm$ 0.21 &  99.831 $\pm$ 0.16 &  -0.220 $\pm$ 0.005   \\
28.1657731 &	-13.9614409  & 316.814 $\pm$ 0.03 & 341.394 $\pm$ 0.03 & 425.256 $\pm$ 0.21 & 552.300 $\pm$ 0.16 &  -0.534 $\pm$ 0.348    \\
28.2023977 &	-13.9609316  &   6.262 $\pm$ 0.03 &   7.842 $\pm$ 0.03 &  12.860 $\pm$ 0.20 &  11.498 $\pm$ 0.16 &  -0.948 $\pm$ 0.524    \\
28.2118833 &	-13.9445033  &   9.623 $\pm$ 0.03 &  10.218 $\pm$ 0.03 &  16.433 $\pm$ 0.20 &  13.709 $\pm$ 0.16 &  -0.429 $\pm$ 0.649    \\
28.2002399 &	-13.9685599  &   4.550 $\pm$ 0.03 &   5.340 $\pm$ 0.03 &   4.764 $\pm$ 0.20 &   5.670 $\pm$ 0.16 &  -0.463 $\pm$ 0.721    \\
28.1785907 &	-13.9401408  &   1.416 $\pm$ 0.03 &   1.910 $\pm$ 0.03 &   3.398 $\pm$ 0.21 &   3.143 $\pm$ 0.16 &  -1.241 $\pm$ 0.727    \\
28.1902671 &	-13.9546800  &  15.310 $\pm$ 0.03 &  17.536 $\pm$ 0.03 &  16.802 $\pm$ 0.21 &  15.846 $\pm$ 0.16 &  -0.313 $\pm$ 0.899    \\
\end{longtable}
\normalsize

\clearpage

\begin{threeparttable}
\small
\tabcolsep=1mm
\begin{longtable}{lrccccccc}
\caption[AGN Overlap]{Multi-wavelength Comparison of AGN Detection Techniques\label{overlap}} \\
\hline
\multicolumn{1}{c}{Cluster} &
\multicolumn{1}{c}{\#Variables} &
\multicolumn{1}{c}{\#IR Power Law} &
\multicolumn{1}{c}{\#IR Power Law} &
\multicolumn{1}{c}{\#X-ray PS} &
\multicolumn{1}{c}{\#X-ray PS} &
\multicolumn{1}{c}{Var$/$IR} &
\multicolumn{1}{c}{X-ray$/$IR} &
\multicolumn{1}{c}{Var$/$X-ray} \\
\multicolumn{1}{c}{} & 
\multicolumn{1}{c}{Total} & 
\multicolumn{1}{c}{Total} & 
\multicolumn{1}{c}{w$/$opt. counterparts} & 
\multicolumn{1}{c}{Total} & 
\multicolumn{1}{c}{w$/$opt. counterparts} & 
\multicolumn{1}{c}{} & 
\multicolumn{1}{c}{} & 
\multicolumn{1}{c}{} \\
\hline
\endhead
$CL0152$          &   12     &    9    &  6    &  4       &   3   &  1\tnote{a}   &      2\tnote{a}	 &	2\tnote{a}     \\
$CLJ1226$        &   20     &    8    &  6    &  6       &   6   &  0   &      1	 &	0     \\
$MACSJ0257$  &   13     &    -     &  -     &  1       &   1   &  -   &      -	 &	0     \\
$MACSJ0717$  &   19     &   18   &  13  &  9        &   7   &  1\tnote{a}   &      1\tnote{a}	 &	3\tnote{a}     \\
$MACSJ0744$  &   18     &    -     &  -     &  12     &   6   &  -   &      -	 &	2     \\
$MACSJ0911$  &   14     &    -     &  -     &  5       &   4   &  -   &      -	 &	1     \\
$MACSJ1149$  &   13     &    -     &  -     &  2       &   2   &  -   &      -	 &	1     \\
$MACSJ1423$  &   24     &    17  &  15  &  10     &   9   &  0   &      0	 &	2     \\
$MACSJ2214$  &   8       & 	 -     &  -     &  1       &   3   &  -   &      -	 &	0     \\
$MS0451$        &   16     &    2    &  1    &  7       &   -    &  0   &      -	 &	-     \\
$MS1054$        &   11     &    7    &  4    &  8       &   3   &  0   &      3	 &	0     \\
$SDSS1004$    &   11     &    3    &  3    &  6       &   6   &  0   &      0	 &	1     \\
\hline
\end{longtable} 
\begin{tablenotes}
\item[a]{One object detected by all 3 techniques.}
\end{tablenotes}
\end{threeparttable}

\clearpage

\begin{table*}
 \centering
 \begin{minipage}{140mm}
\caption[Galaxies in the Lacy Wedge]{Galaxies in the Lacy Wedge\label{lacywedge}}
\begin{tabular}{lcccc}
\hline
\multicolumn{1}{c}{Cluster} &
\multicolumn{1}{c}{Galaxies} &
\multicolumn{1}{c}{IR Power Law} &
\multicolumn{1}{c}{Optical Variables} &
\multicolumn{1}{c}{X-ray Point Sources} \\
\hline
$CL0152$        &   65/267  & 9/9      & 2/5    & 4/4  \\
$CLJ1226$      &   90/332  & 8/8      & 3/16  & 2/6  \\
$MACSJ0717$ &   49/288  & 15/18  & 2/8   & 2/5  \\
$MACSJ1423$ &   55/247  & 13/17  & 2/12 & 1/8  \\
$MS0451$       &   42/293  & 2/2      & 0/8    & 2/6  \\
$MS1054$       &   54/319  & 7/7      & 0/6    & 4/6  \\
$SDSS1004$   &   56/251  & 3/3      & 1/9    & 3/6  \\
\hline
\end{tabular}
\end{minipage}
\end{table*}

\begin{threeparttable}
\small
\begin{longtable}{lccccc}
\caption[Cluster Spectroscopic Coverage]{Cluster Spectroscopic Coverage \label{zsandranges}} \\
$SDSS1004$  &    3    &  3    &  -   &  0.68    &   - \kill
\hline
\multicolumn{1}{c}{Cluster} &
\multicolumn{1}{c}{\# Redshifts} &
\multicolumn{1}{c}{\# Members} &
\multicolumn{1}{c}{\# Field} &
\multicolumn{1}{c}{Redshift} & 
\multicolumn{1}{c}{Cluster Redshift Range\tnote{*}} \\
\hline
\endhead
$CL0152$\tnote{a}         &   81    &  48   &  33  &  0.831   &   0.82 - 0.87 \\
$CLJ1226$\tnote{b}       &   42    &  25   &  17  &  0.888   &   0.90 - 0.91  \\
$MACSJ0257$\tnote{c}  &   16    &  13   &  3   &  0.506   &   0.50 - 0.51 \\
$MACSJ0717$\tnote{c}  &   125   &  108  &  17  &  0.548   &   0.53 - 0.56 \\
$MACSJ0744$\tnote{c}  &   34    &  31   &  3   &  0.686   &   0.68 - 0.71 \\
$MACSJ0911$\tnote{c}  &   24    &  21   &  3   &  0.504   &   0.49 - 0.52 \\
$MACSJ1149$\tnote{c}  &   44    &  42   &  2   &  0.544   &   0.52 - 0.56 \\
$MACSJ1423$\tnote{c}  &   45    &  36   &  9   &  0.545   &   0.53 - 0.56 \\
$MACSJ2214$\tnote{c}  &   55    &  52   &  3   &  0.504   &   0.49 - 0.52 \\
$MS0451$\tnote{d}        &   127   &  68   &  59  &  0.550    &   0.52 - 0.56 \\
$MS1054$\tnote{e}        &   206   & 108   &  98  &  0.830    &   0.80 - 0.86 \\
$SDSS1004$                                &    3    &  3    &  -   &  0.680    &   - \\
\hline
\end{longtable}
\begin{tablenotes}
\item[*]{Lowest-highest redshifts of confirmed cluster members from the literature}
\item[a]{\citet{demarco05}}
\item[b]{\citet{ellis06}}
\item[c]{\citet{barrett06}}
\item[d]{\citet{moran07}}
\item[e]{\citet{tran07}, private communication}
\end{tablenotes}
\end{threeparttable}

\clearpage

\begin{table*}
 \centering
 \begin{minipage}{140mm}
\caption[AGN Redshifts]{AGN Redshifts \label{agnzs} \footnote{A dash (-) indicates that data is not available.}}
\begin{tabular}{lcccccc}
\hline
\multicolumn{1}{c}{Cluster} &
\multicolumn{1}{c}{Var} &
\multicolumn{1}{c}{Var} &
\multicolumn{1}{c}{IR} &
\multicolumn{1}{c}{IR} &
\multicolumn{1}{c}{X-ray} &
\multicolumn{1}{c}{X-ray} \\
\multicolumn{1}{c}{} & 
\multicolumn{1}{c}{Cluster} &
\multicolumn{1}{c}{Field} &
\multicolumn{1}{c}{Cluster} &
\multicolumn{1}{c}{Field} &
\multicolumn{1}{c}{Cluster} &
\multicolumn{1}{c}{Field} \\
\hline
$CL0152$         &    1    &    1	  &     2   &	 0    &   2   &  1    \\
$CLJ1226$       &    1    &    0	  &     0   &	 0    &   1   &  0    \\
$MACSJ0257$  &    0    &    0	  &     -    &	 -    &   1   &  0    \\
$MACSJ0717$  &    1    &    1	  &     0   &	 0    &   2   &  2    \\
$MACSJ0744$  &    1    &    0	  &     -    &	 -    &   1   &  0    \\
$MACSJ0911$  &    0    &    1	  &     -    &	 -    &   0   &  0    \\
$MACSJ1149$  &    1    &    0	  &     -    &	 -    &   0   &  0    \\
$MACSJ1423$  &    5    &    2	  &     0   &	 0    &   2   &  2    \\
$MACSJ2214$  &    0    &    0	  &     -    &	 -    &   1   &  0    \\
$MS0451$        &    3    &    2	  &     1   &	 0    &   -   &  -    \\
$MS1054$        &    2    &    2	  &     1   &	 1    &   1   &  0    \\
$SDSS1004$    &    -    &    -	  &     -    &	 -    &   -   &  -    \\
\hline
\end{tabular}
\end{minipage}
\end{table*}

\begin{table*}
 \centering
 \begin{minipage}{140mm}
\caption[Percentage of Cluster AGN]{Percentage of Cluster AGN \label{AGNpercenttable} \footnote{A dash (-) indicates that data is not available; an x indicates there are no AGN detected, though data exist.}}
\begin{tabular}{lccc}
\hline
\multicolumn{1}{c}{Cluster} &
\multicolumn{1}{c}{\% AGN$_{weighted}$} &
\multicolumn{1}{c}{\% AGN$_{non-weighted}$} &
\multicolumn{1}{c}{\% AGN$_{spec}$} \\
\hline
$CL0152$         & 2.49 (21/889)     &   2.36 (21/889)     & 	4.17 (2/48)    \\
$CLJ1226$       & 3.13 (33/961)     &   3.43 (33/961)     & 	8.0 (2/25)      \\
$MACSJ0257$  & 1.0 (14/1535)     &   0.91 (14/1535)   & 	7.7 (1/13)      \\
$MACSJ0717$  & 2.27 (42/1813)   &   2.32 (42/1813)   & 	2.78 (3/108)  \\
$MACSJ0744$  & 2.13 (28/1430)   &   1.96 (28/1430)   & 	3.23 (1/31)    \\
$MACSJ0911$  & 1.35 (18/1353)   &   1.33 (18/1353)   & 	x                    \\
$MACSJ1149$  & 0.92 (13/1512)   &   0.86 (13/1512)   & 	2.38 (1/42)    \\
$MACSJ1423$  & 3.75 (49/1337)   &   3.66 (49/1337)   & 	20.0 (7/35)    \\
$MACSJ2214$  & 0.81 (12/1373)   &   0.87 (12/1373)   & 	1.92 (1/52)    \\
$MS0451$        & 2.65 (25/1102)   &   2.27 (25/1102)   & 	5.88 (4/68)    \\
$MS1054$        & 2.29 (23/893)     &   2.58 (23/893)     & 	2.78 (3/108)  \\
$SDSS1004$    & 1.43 (20/1651)   &   1.21 (20/1651)   & 	-	             \\
\hline
Mean               & 2.01 $\pm$ 1.4   &   1.98 $\pm$ 1.4  &        5.88 $\pm$ 2.3 \\
Median            & 2.27 $\pm$ 1.5   &    2.27 $\pm$ 1.5 &        4.17 $\pm$ 2.0 \\
\hline
\end{tabular}
\end{minipage}
\end{table*}

\clearpage

\begin{longtable}{lccc}
\caption[Percentage of Cluster AGN By Type]{Percentage of Cluster AGN By Type\label{AGNpercenttablebytype}} \\
$CL0152$          & 1.16 (9/889)     &  4.17 (2/48)  \kill
\hline
\multicolumn{1}{c}{Cluster} & 
\multicolumn{1}{c}{\% AGN$_{weighted}$} & 
\multicolumn{1}{c}{\% AGN$_{spec}$} \\
\hline
\endfirsthead
{\tablename\ \thetable{}. Continued} \\
\hline
\multicolumn{1}{c}{Cluster} & 
\multicolumn{1}{c}{\% AGN$_{weighted}$} & 
\multicolumn{1}{c}{\% AGN$_{spec}$} \\
\hline
\endhead
\hline
\endfoot
\hline
\endlastfoot
\multicolumn{1}{l}{\bf Variables} \\
$CL0152$        & 1.37 (12/889)   & 2.08 (1/48)  \\
$CLJ1226$      & 2.05 (20/961)   & 4.0 (1/25)   \\
$MACSJ0257$ & 0.92 (13/1535) & -            \\
$MACSJ0717$ & 1.04 (19/1813) & 0.93 (1/108) \\
$MACSJ0744$ & 1.38 (18/1430) & 3.23 (1/31)  \\
$MACSJ0911$ & 1.05 (14/1353) & x            \\
$MACSJ1149$ & 0.86 (12/1512) & 2.38 (1/42)  \\
\end{longtable}

\label{lastpage}

\end{document}